%% file: main.tex
\documentclass[sigconf]{acmart}

\usepackage{pifont}
\usepackage{enumitem}
\usepackage{xspace}
\usepackage{threeparttable}
\usepackage{tabularx}  %
\usepackage{ragged2e}
\usepackage{array}     %
\usepackage{amsthm, thmtools}
\usepackage{amsthm, thmtools}
\usepackage{graphicx}
\usepackage{subcaption}
\usepackage{arydshln}
\usepackage{makecell, cellspace}
\usepackage{multirow}
\usepackage{threeparttable}

\newcommand{\appname}{{\textsc{PatchDiff}}\xspace}

\AtBeginDocument{%
  }

\copyrightyear{2026}
\acmYear{2026}
\setcopyright{rightsretained}
\acmConference[ICSE '26]{2026 IEEE/ACM 48th International Conference on Software Engineering}{April 12--18, 2026}{Rio de Janeiro, Brazil}
\acmBooktitle{2026 IEEE/ACM 48th International Conference on Software Engineering (ICSE '26), April 12--18, 2026, Rio de Janeiro, Brazil}
\acmPrice{}
\acmDOI{10.1145/3744916.3764576}
\acmISBN{979-8-4007-2025-3/26/04}

\begin{document}

\title[Are ``Solved Issues'' in SWE-bench Really Solved Correctly?]{Are ``Solved Issues'' in SWE-bench Really Solved Correctly? \\An Empirical Study}

\author{You Wang}
\affiliation{%
\institution{The State Key Laboratory of Blockchain and Data Security, Zhejiang University}
\city{Hangzhou}
\country{China}}
\email{prinzywang@zju.edu.cn}

\author{Michael Pradel}
\affiliation{%
  \institution{CISPA Helmholtz Center for Information Security}
  \city{Stuttgart}
  \country{Germany}}
\email{michael@binaervarianz.de}

\author{Zhongxin Liu}
\authornote{Corresponding Author}
\affiliation{%
  \institution{The State Key Laboratory of Blockchain and Data Security, Zhejiang University}
  \city{Hangzhou}
\country{China}}
\email{liu_zx@zju.edu.cn}

\begin{abstract}

Automated issue solving aims to resolve real-world issues in software repositories.
The most popular benchmarks for automated issue solving are SWE-bench and its human-filtered subset SWE-bench Verified, which are widely used to evaluate foundation models and software engineering agents.
These benchmarks leverage testing to validate generated patches.
However, because testing is rarely exhaustive, a patch may pass the tests but nevertheless fail to match the developers' expectations.
Unfortunately, it is currently unclear to what extent evaluations performed with SWE-bench suffer from such \emph{plausible but incorrect patches}.
This paper presents an in-depth empirical study of the correctness of plausible patches generated by three state-of-the-art issue-solving tools (CodeStory, LearnByInteract, and OpenHands) evaluated on SWE-bench Verified.
We extensively test and inspect generated patches, and compare them against human-written ground truth patches.
The core of our methodology is a novel technique for differential patch testing, called \appname, which automatically exposes behavioral discrepancies between two patches.
Our findings reveal critical weaknesses in SWE-bench's patch validation mechanism, which causes 7.8\% of all patches to count as ``correct'' while failing the developer-written test suite.
Moreover, our novel automated technique reveals that even more (29.6\%) plausible patches induce different behavior than the ground truth patches.
These behavioral differences are often due to similar, but divergent implementations (46.8\%) and due to generated patches that adapt more behavior than the ground truth patches (27.3\%).
Our manual inspection shows that 28.6\% of behaviorally divergent patches are certainly incorrect.
Combined, the different weaknesses lead to an inflation of reported resolution rates by 6.4 absolute percent points.
Our findings are a call to arms for more robust and reliable evaluation of issue-solving tools.
We envision our automated differential patch testing technique to be useful for this purpose.
\end{abstract}



\pagestyle{plain}

\maketitle

\input{tex/introduction}

\input{tex/preliminary}

\input{tex/approach}

\input{tex/empirical}

\input{tex/discussion}

\input{tex/related_work}
\input{tex/conclusion}

\input{tex/data}
\input{tex/acks}

\bibliographystyle{ACM-Reference-Format}
\balance
\bibliography{references,referencesMichael}

\end{document}

%% file: tex/introduction.tex
\section{Introduction}~\label{sec:intro}

Automated issue solving aims to address real-world issues in software repositories, and holds great potential to reduce maintenance costs and improve software quality.
SWE-bench~\cite{jimenez2023swe} stands out as the most popular benchmark for automated issue solving, comprising 2,294 tasks from 12 well-maintained Python repositories.
Each task is provided with an issue statement describing the specific task to accomplish and the repository version where the issue is to be solved.
The tool to be evaluated is then required to generate a patch to address the issue. 
SWE-bench evaluates the generated patch by running tests associated with the issue, including at least one fail-to-pass test to confirm the issue is resolved, and, where available, several pass-to-pass tests to ensure no regression in existing functionality.
Recently, OpenAI hired 93 developers to manually identify and filter out those SWE-bench tasks with overly specific and even issue-unrelated tests, and curated a subset comprising 500 tasks, known as SWE-bench Verified~\cite{swebenchverified}.
The SWE-bench leaderboard is getting immense attention.
It has been used to judge the merit of new techniques proposed in academic papers~\cite{zhang2024autocoderover,xia2024agentless}, and also to evaluate commercial tools for assessing their potential value~\cite{wandbai, isoformai, blackboxai}.
Moreover, SWE-bench Verified has been widely used to evaluate the coding abilities of state-of-the-art foundation models, such as OpenAI GPT-o1~\cite{gpt-o1} and Anthropic Claude-3.5~\cite{claude-3.5}.

However, due to practical limitations, test suites are rarely exhaustive and often suffer from weaknesses~\cite{smith2015cure}.
Following the literature on automated program repair~\cite{liu2019tbar}, we refer to patches that pass their corresponding validation process as \emph{plausible patches}.
Validation with weak test suites can result in \emph{plausible but incorrect patches}, inflating the performance of the evaluated tools, and possibly leading to incorrect conclusions about their abilities.
Although others have noticed the problem of weak test suites in SWE-bench~\cite{swebenchverified, aleithan2024swe}, it is unclear to what extent evaluations performed with SWE-bench suffer from this problem, and there is a notable lack of effective methodologies for detecting plausible but incorrect patches.

This paper conducts an in-depth empirical study of the correctness of plausible generated patches on SWE-bench.
We focus on the high-quality, human-filtered, and widely used subset SWE-bench Verified, and conduct this study on the plausible patches generated by the state-of-the-art issue-solving tools, i.e., CodeStory~\cite{codestory}, LearnByInteract~\cite{su2025learn}, and OpenHands~\cite{wang2024openhands}.
Our study addresses four research questions:

\noindent \textbf{RQ1: What is the impact of executing all developer tests in SWE-bench?}
We first analyze the validation process of SWE-bench and identify a flaw that weakens the test suites.
Specifically, when validating a generated patch, SWE-bench only uses the developer-written test files modified in the pull request (PR) for fixing the target issue, potentially leaving functionality covered by other test files untested. 
To quantify the impact of this flaw,
we execute all available test files in the corresponding repository to re-validate each patch.
The results show that, on average, 7.8\% of plausible patches are incorrect, leading to an absolute drop of the issue resolution rate of 4.5\%, on average. 
This indicates that neglecting the test files not modified in the PR weakens the test suite and can inflate reported performance.

The findings from RQ1 further raise two critical questions:
Are there any patches that pass all developer tests but remain incorrect? 
If yes, how many?
Answering these questions is non-trivial.
One option is to manually inspect generated plausible patches and compare them with their corresponding developer-written ground truth patch (hereon, \emph{oracle patches}).
However, this approach is labor-intensive, error-prone, and does not scale.
Another option is to generate more regression tests based on the fixed repository version to strengthen the validation~\cite{xin2017identifying,yu2019alleviating}.
However, this approach typically generates plenty of test cases with most of them unrelated to the generated patch, suffering from limited effectiveness.
Moreover, according to our investigation (Section~\ref{subsec:comparison}), there is a lack of robust and effective test generation tools for real-world Python projects.

To enable this study, we present a novel differential patch testing technique, named \appname, which aims to generate test cases that expose behavioral discrepancies between the plausible patch and the developer-written oracle patch.
We refer to such generated tests as \emph{differentiating tests}.
\appname leverages Large Language Models (LLMs) for test generation and is enabled by a call-trace-based method to identify appropriate target functions and construct useful contextual information for LLMs.
Specifically, we first leverage \appname to generate differentiating tests for each plausible patch and identify behaviorally divergent patches, and then manually assess the correctness of these suspicious patches with the help of the differentiating tests.
This approach provides concrete evidence of invalid functionality, enables more focused and objective patch validation, and reduces manual validation effort.

With the help of \appname, we address the following research questions: 

\noindent \textbf{RQ2: How many generated plausible patches exhibit behavioral discrepancies compared to their oracle patches?}
To answer this question, we leverage \appname to generate differentiating tests for the plausible patches assessed in RQ1.
Our findings reveal that, on average, 29.6\% of plausible patches can be differentiated from their oracle patches through the tests generated by \appname.
We refer to such patches as \emph{suspicious patches}.
This highlights that a substantial proportion of plausible patches are likely to diverge from the expected behavior, raising concerns about their correctness.

\noindent \textbf{RQ3: What are the patterns of differences between plausible and oracle patches that lead to behavioral discrepancies?}
Understanding these patterns offers valuable insights into how suspicious patches deviate from their oracle patches and can inspire the development of more human-aligned issue-solving tools.
To answer this question, 
we sample 77 (30\%) suspicious patches from those identified in RQ2, manually compare them with their oracle patches, and craft a taxonomy of the patch differences leading to behavioral discrepancies.
Our analysis reveals that behavioral discrepancies between plausible and oracle patches are often due to similar but divergent implementations (46.8\%) and due to plausible patches adapting more behavior than their oracle patches (27.3\%)

\noindent \textbf{RQ4: In cases of behavioral discrepancies, how many generated plausible patches are incorrect?}
A behavioral discrepancy reveals incorrectness only when the behavior of the generated patch violates the expected behavior. 
Therefore, the correctness of a suspicious patch requires further validation.
To answer this question, 
we manually assess the correctness of each suspicious patch evaluated in RQ3 based on its differentiating tests, developer-written tests, the repository, the issue statement, and the oracle patch. 
We find that 28.6\% of suspicious patches are certainly incorrect.
If we assume that incorrect patches are distributed evenly among suspicious patches, this result extrapolates to an approximated incorrectness rate of 11.0\% among plausible patches, which inflates the resolution rates of the studied tools by 6.4 absolute percent points, on average.
This result further raises concerns about the reliability of SWE-bench's validation mechanism. 

Our findings provide actionable insights for users and maintainers of issue-solving benchmarks, such as carefully selecting developer tests for patch validation, checking and filtering out plausible but incorrect patches for more accurate evaluation, and paying more attention to supplementary semantic changes in plausible patches.
In addition, the under-specified issue statements in SWE-bench Verified call for better issue-solving tools that are capable of detecting and refining vague specifications and better issue-solving benchmarks where issues are well specified.
We also envision \appname to be useful for sustainably strengthening issue-solving benchmarks.
Specifically, practitioners can use the differentiating tests generated by \appname to ease their check for incorrect patches.
The generated tests that successfully identify plausible but incorrect patches can be incorporated into the test suites in benchmarks.
Over time, as test suites continue to evolve and become comprehensive, fewer manual efforts are required and a sustainable and robust patch validation ecosystem for issue-solving tools can be built.

In summary, the main contributions of this paper are as follows:

\vspace{-2pt}

\begin{itemize}[leftmargin=*]

    \item \emph{In-depth study}. The first in-depth study on the correctness of generated plausible patches on SWE-bench. 

    \item \emph{Technique}. A novel differential patch testing technique \appname that can generate tests to reveal meaningful behavioral differences between patches.

    \item \emph{Insights}. Insights for users and maintainers of issue-solving benchmarks towards more robust and sustainable evaluation.

    \item \emph{Replication package}. A replication package~\cite{package} including the implementation of \appname and the results of our study.

\end{itemize}

%% file: tex/preliminary.tex
\section{Background and Motivating Example}~\label{sec:background}
This section first introduces SWE-bench and SWE-bench Verified and then describes a motivating example.

\subsection{SWE-bench and SWE-bench Verified}
SWE-bench~\cite{jimenez2023swe} is the most popular benchmark for automated issue solving.
As described in Section~\ref{sec:intro}, to assess the effectiveness of an issue-solving tool on SWE-bench, the user needs to generate patches for each task based on the corresponding issue statement and the buggy repository version.
SWE-bench also provides a test patch for each task, which includes the changes made to test files in the PR that resolves the corresponding issue.
The full set of SWE-bench is shown to consist of low-quality and noise instances~\cite{swebenchverified}.
OpenAI curated a high-quality subset known as SWE-bench Verified, which consists of 500 samples.

To validate the correctness of a generated patch, both SWE-bench and SWE-bench Verified run the test files modified in the test patch after applying both the test patch and the generated patch to the buggy repository version. 
If these test files all pass, the patch is regarded as correct~\cite{harness-swebench}.
Although such test files are likely to be relevant to the issue, they do not necessarily cover all the functionalities that can be affected by the generated patches.
As a result, this validation process is weak and may accept incorrect patches that violate the uncovered functionalities.

\subsection{Motivating Example} \label{subsec:example}

\begin{figure}
    \centering
    \includegraphics[width=1.0\linewidth]{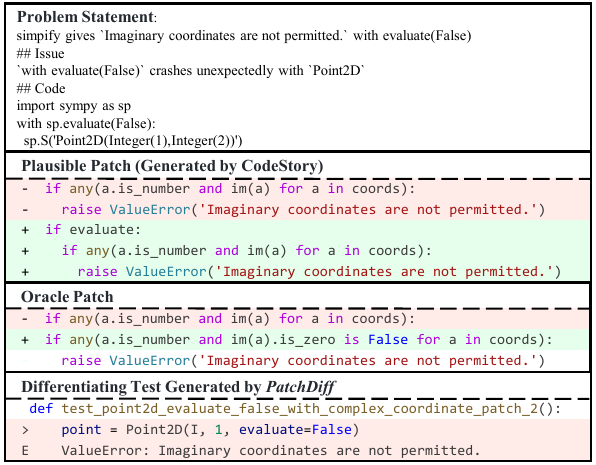}
    \vspace{-0.6 cm}
    \caption{An example of plausible but incorrect patches from the issue sympy-22714, where the exception is correctly raised only under the oracle patch}
    \label{fig:rq4_example}
    \vspace{-0.3 cm}
\end{figure}

Figure ~\ref{fig:rq4_example} presents an example of a plausible but incorrect patch generated by CodeStory.
This patch attempts to address the issue that, when a \verb|Point2D| object is created under \verb|evaluate(False)|, 
the program incorrectly raises a \verb|ValueError| with the message "Imaginary coordinates are not permitted", even though there are no imaginary inputs. 
To assess the correctness of this patch, we manually compare the implementation of the two patches. 
The generated plausible patch modifies the safety-checking code by introducing a condition \verb|if evaluate:|,  
and the oracle patch replaces \verb|im(a)| with \verb|im(a).is_zero()|. 
However, this does not directly give us enough information to determine the correctness of the generated patch. 
We have to read the definitions of \verb|im()| and \verb|is_zero()| in the repository to fully understand the issue. 
In fact, the underlying bug is caused by the method \verb|im()| consistently returning a truthy value when \verb|evaluate| is \verb|False|, regardless of the input. 
The oracle patch resolves the issue by explicitly invoking \verb|im().is_zero()| to determine the presence of imaginary numbers, 
while the generated patch simply mutes the examination of imaginary inputs when \verb|evaluate| is \verb|False|, 
allowing an invalid object creation to proceed. 
This suggests that the generated patch fails to correctly locate and fix the bug, and thus should be an incorrect patch. 

The weak test suite used by SWE-bench for this issue leads to such plausible but incorrect, which leads to performance overestimation and can cause misleading comparisons between tools.
Manually identifying such plausible but incorrect patches is challenging and labor-intensive, and the manual efforts are not reusable.
That is, another manual review is required if another similar but not identical plausible patch is generated.
Providing tests that can expose meaningful behavioral differences between the plausible and oracle patches can significantly ease the assessment of patch correctness.
In the above example, 
the differentiating test generated by \appname intentionally attempts to create a \verb|Point2D| object with imaginary coordinates while setting \verb|evaluate| to \verb|False|.
The oracle patch correctly triggers an exception, preventing the creation of the invalid object, 
whereas the suspicious patch fails to trigger this error.
This test directly demonstrates that the generated patch is incorrect, reducing manual efforts.
Moreover, it can be added to the test suite to strengthen the patch validation of this issue.

%% file: tex/approach.tex
\section{Methodology}
This section describes the methodology of our study, including the issue solving tools and patches we use and our differential patch testing technique.

\subsection{Issue Solving Tools and Patches}
Our empirical study focuses on SWE-bench Verified, a high-quality, human-validated subset of SWE-bench.
The study is conducted on the plausible patches generated by three state-of-the-art issue-solving tools, i.e., CodeStory Midwit Agent + swe-search~\cite{codestory} (hereinafter, CodeStory), Learn-by-interact~\cite{su2025learn} (hereinafter, LearnByInteract), and OpenHands + CodeAct v2.1~\cite{wang2024openhands} (hereinafter, OpenHands), on SWE-bench Verified.
These tools are selected because they open source their implementations and/or release their papers or technical reports, making them suitable for further research.
Note that for each tool and each issue in SWE-bench Verified, there is at most one plausible patch.

\subsection{Differential Patch Testing}
To enable this study, we proposes an automated differential patch testing technique named \appname.
Given an issue-solving task providing the issue statement and the buggy repository version, its test patch $P_t$, its oracle patch $P_o$, and a patch $P_g$ generated for it, \appname leverages an LLM to generate a set of tests that can reveal the behavioral differences between $P_g$ and $P_o$, i.e., \emph{differentiating tests}.
We refer to the repository version with only $P_t$ applied as $R_t$ and with both $P_t$ and $P_g$ or $P_o$ applied as $R_g$ or $R_o$, respectively.
\appname first checks whether $P_g$ and $P_o$ are syntactically identical without considering comments, and omits identical $P_g$.
Then, \appname leverages a call-trace-based method to identify appropriate target functions and extract contextual code.
Finally, \appname prompts an LLM to generate and repair tests for target functions, and filters out unqualified tests.

\subsubsection{Target Function Identification}~\label{subsubsec:target-func}
To generate tests, \appname first needs to determine the target function to be tested.
It regards a function that satisfies the following criteria as a target function: (1) The function is a patch-modified function or a function that directly or indirectly invokes a patch-modified function. (2) The function is not defined within test files. (3) The function is directly invoked by developer-written tests.
Such a function is related to patches and not related to tests, and developers usually have specific expectations regarding its behavior.
To identify target functions, we first instrument each patch-modified function in $R_g$ and $R_o$, and run all test files in $R_g$ and $R_o$ to collect call traces.
Each call trace starts from a test function in $R_g$ or $R_o$ and ends at a patch-modified function.
The first non-test function in each call trace satisfies the criterion mentioned above and \appname annotates it as a target function.
Different call traces can have the same target function.
So a target function may correspond to multiple call traces.

\subsubsection{Contextual Code Extraction}
In each call trace, the functions in test files provide information about how to invoke the target function, and the functions in non-test files are either the target function or illustrate how the target function utilizes the patch-modified function.
So all the functions in the call trace are useful for generating differentiating tests.
For each target function, we extract the functions from the shortest call trace collected in $R_g$ and the shortest one collected in $R_o$.
For each of these functions, we further map it to its before-patch version in $R_t$ based on its class and function names.
The mapped functions provide the contextual information before patches are applied and are referred to as \emph{context functions}.
Only providing function definitions to the LLM may miss critical contextual information such as class information.
Therefore, we extend context functions to construct contextual code.
Specifically, we collect all Python files that contain the target function or at least one context/patch-modified function from $R_t$. We remove the functions that are not target, context, or patch-modified functions from these files.
Any classes rendered empty after this deletion are also discarded.
In addition, in test files, we annotate code lines where the target function is invoked with a comment to direct
the LLM’s attention.
All the streamlined files form the contextual code of this target function.

\subsubsection{LLM-Based Test Generation}
We prompt the LLM to generate differentiating tests for one target function at a time.
It is costly to generate tests for many target functions with LLMs.
Thus, we select at most 10 target functions for test generation. In detail, we calculate the number of non-test functions $l$ in each collected call trace. For each target function, we assign its smallest $l$ as its score.
The 10 target functions with the smallest scores are selected.
The rationale behind this selection is that a smaller $l$ indicates a simpler relationship between the patch-modified function and the target function, easing the LLM to discern and trigger behavioral discrepancies.

For each of the selected target functions, we construct a prompt by incorporating itself, $P_o$, $P_g$, its contextual code, and its shortest call traces obtained from $R_o$ and $R_g$, respectively. 
The two example call traces demonstrate how a developer test exercises the target function, and how the target function eventually invokes the patch-affected functions. 
In the prompt, we instruct the LLM to first compare the two patches and reason how the patch-modified functions affect the target function through a chain-of-thought analysis, and then generate a new test file that (1) specifically tests the target function and (2) passes under one patch but fails under another patch. 
For each target function, we request the LLM to generate 10 responses in one request. 

For each generated test file, if it fails to differentiate the patches and some tests in it fail under both patches, 
we further prompt the LLM with this test file and the test results under $P_o$, and instruct it to repair the tests so that they get passed on $R_o$. 
This step aims to repair the tests with erroneous implementations and the assertions with expected output different from the output of $R_o$.
The repairing process iterates for 2 cycles. 

To balance costs and effectiveness, we employ the OpenAI gpt-4o-mini-2024-07-18 model as the underlying LLM for test generation. 
This model is configured with a temperature setting of 1, allowing the generation of diverse outputs that enhance the likelihood of identifying meaningful behavioral discrepancies.

\subsubsection{Unqualified Test Filtering}
\appname aims at generating tests for target functions to expose meaningful behavioral discrepancies.
However, since LLMs do not always follow the instructions, 
the generated tests may not only examine the specified target function.
To increase the probability of exposing meaningful behavioral discrepancies, \appname filters out such tests.
Specifically, we instrument patch-modified functions and execute each generated differentiating test to collect call traces.
For each differentiating test, if there is a call trace where the function directly invoked by the test function is not a target function, we filter it out.
Because this test does not only examine target functions.
We further filter out flaky tests.
In detail, for each remaining differentiating test, we run it under $P_g$ and $P_o$ for 20 times each. 
If it passes under one patch for all 20 times and fails under the other patch for at least one time, we regard it as valid.
Otherwise, we filter out it.
All the remaining differentiating tests are regarded as the output of \appname.

%% file: tex/empirical.tex
\section{Empirical Study}

\input{tex/rq1}

\input{tex/rq2}

\subsection{RQ3: Patterns of Patch Differences Leading to Behavioral Discrepancies}
Behavioral discrepancies revealed by generated differentiating tests provide evidence for assessing the correctness of plausible patches.
This RQ aims to investigate the patterns of the differences between patches that lead to behavioral discrepancies.
The answer to this RQ can help us better understand the cause of the observed behavioral discrepancies and 
provide insights into developing more robust issue-solving tools. 

\noindent \textbf{Approach:}
We sample 77 (30\%) of the suspicious patches detected in RQ2 for investigation and manually derive a taxonomy of patch difference patterns leading to behavioral discrepancies.

\begin{figure*}
    \centering
    \includegraphics[width=0.95\linewidth]{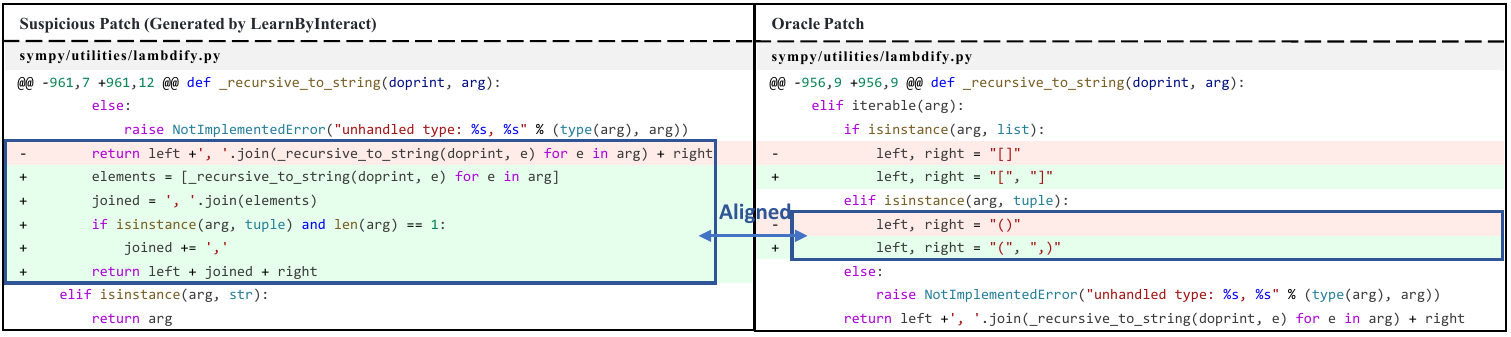}
    \vspace{-0.3 cm}
    \caption{Example of aligned sem-changes (sympy-23262)}
    \label{fig:rq3_example_atomic}
    \vspace{-0.3 cm}
\end{figure*}

To analyze such patch difference patterns, we need to identify the commonalities and differences between each suspicious patch and its oracle patch.
However, this comparison is non-trivial, because the two patches tend to have divergent syntactical implementations.
To address this, we introduce the concept of atomic semantic changes (called, \emph{sem-changes} for brevity).
Sem-changes refer to a set of code line edits in a patch that alter program execution behaviors by adding or modifying a specific behavior
(e.g., handling errors under certain conditions, processing specific input types or special cases, or implementing the algorithm to compute a value), 
as opposed to purely structural or syntactical changes. 
These sem-changes directly impact whether and how the patch addresses the requirements of the targeted issue. 
Sem-changes that add or modify the same behavior (maybe in different ways and may not be semantically equivalent) are manually aligned between the suspicious patch and the oracle patch, providing a foundation for identifying the key patch differences that lead to the observed behavioral discrepancies. 
Figure \ref{fig:rq3_example_atomic} illustrates an example of aligned sem-changes. 
In this case, the two diff hunks in the generated and oracle patches form two sem-changes that modify the same behavior, i.e., when the variable \verb|arg| is of type \verb|tuple|, an extra comma is added before the closing parenthesis at the end of the returned string. 
Thus they are aligned although these edits are syntactically different and occur at different positions.

The first author (A1) performs the labeling process.
For any case with uncertain categories, A1 discusses with another author (A2) and reaches a consensus as the final result.
Specifically, for each suspicious patch, we first read it and its corresponding oracle patch, making our best effort to understand how this two patches try to solve the issue based on their context in the repository.  
Then, we extract sem-changes in the two patches that add or modify the same program behavior and align them.
It is possible that a sem-change in one patch does not align with any sem-change in another, i.e., it is unaligned. 
If there is no alignment established between the patches, the patch difference pattern is classified as \ding{182} 
\textit{No Alignment}.
Otherwise, we further review the differentiating tests of this suspicious patch and the different testing outputs under the two patches, 
and then determines which sem-change directly leads to the observed behavioral discrepancies.
We refer to such change as \emph{root change}.
Based on the root change, the patch difference is classified into one of the three categories: \ding{183} \textit{Supplementary Sem-Change}, \ding{184} \textit{Absent Sem-Change}, and  \ding{185} \textit{Divergent Implementations of Sem-Change}.
We further manually analyze the patches under the category of Supplementary Sem-changes, 
and classify them into subcategories based on the program behavior that the root change is adding or modifying.
Theoretically, there can be cases where the observed behavioral discrepancies are attributed to hybrid patch difference patterns.
However, we do not observe such cases in the sampled suspicious patches, 
possibly because the generated plausible patches typically correspond to issues that are not very complex. 
Therefore, we exclude hybrid patterns from our discussion.

\begin{figure*}[t]
    \centering
    
    \begin{subfigure}[t]{\textwidth}
        \centering
        \includegraphics[width=0.95\linewidth]{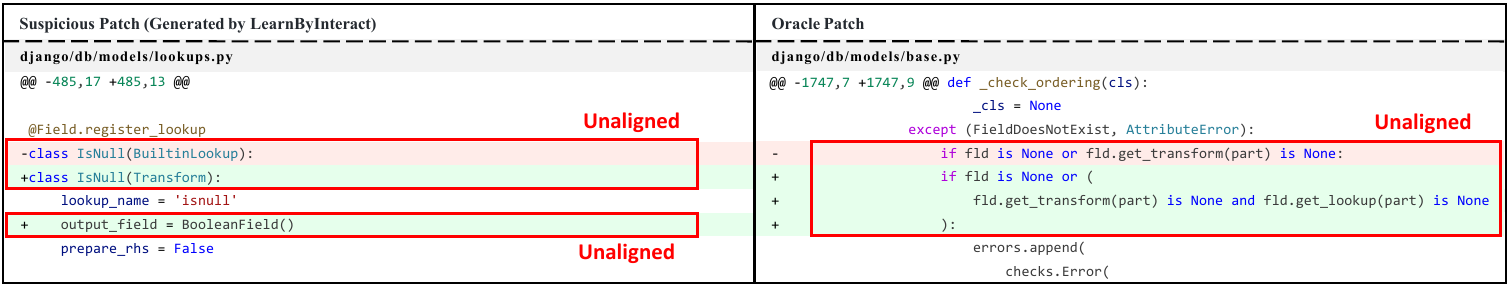}
        \vspace{-0.1cm}
        \subcaption{No alignment (django-12858)}
        \label{fig:rq3_example_very_different}
    \end{subfigure}
    
    \vspace{0.2cm}  %
    
    \begin{subfigure}[t]{\textwidth}
        \centering
        \includegraphics[width=0.95\linewidth]{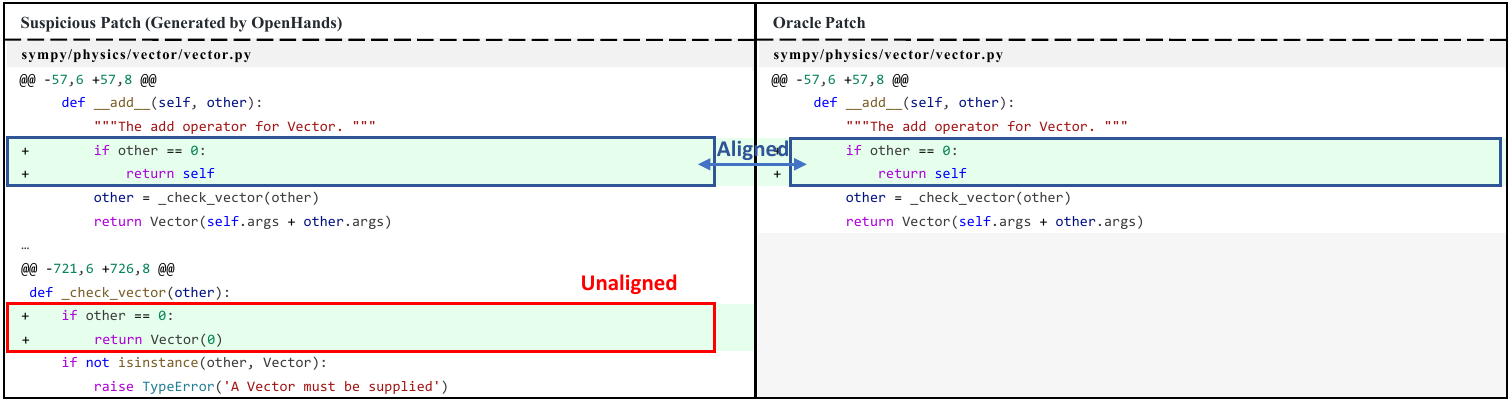}
        \vspace{-0.1cm}
        \subcaption{Supplementary sem-change (sympy-14711)}
        \label{fig:rq3_supplementary}
    \end{subfigure}
    
    \vspace{0.2cm}  %
    
    \begin{subfigure}[t]{\textwidth}
        \centering
        \includegraphics[width=0.95\linewidth]{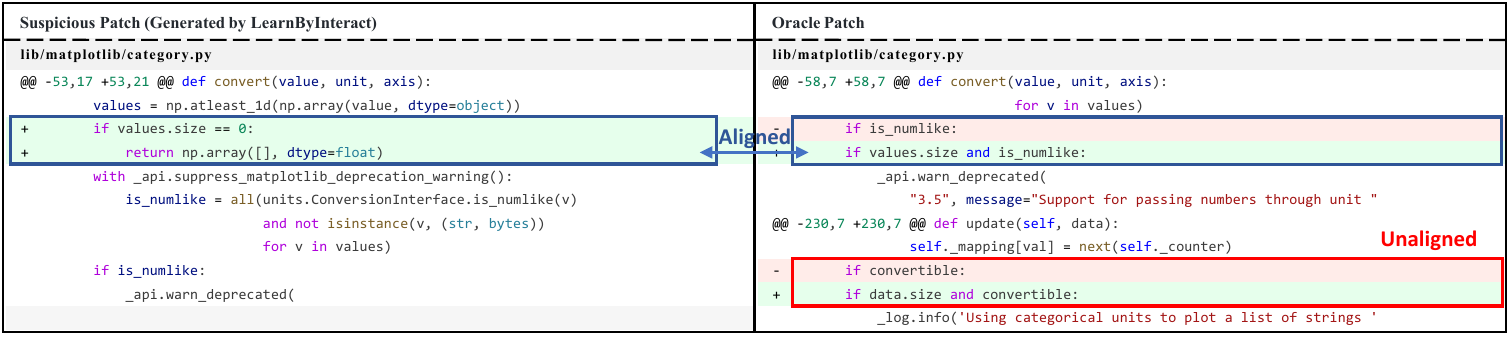}
        \vspace{-0.1cm}
        \subcaption{Absent sem-change (matplotlib-22719)}
        \label{fig:rq3_absent}
    \end{subfigure}
    
    \vspace{-0.3cm}
    \caption{Three examples of patch difference patterns}
    \label{fig:rq3_examples}
    \vspace{-0.3cm}
\end{figure*}

\input{tex/tables/rq3_taxonomy}

\noindent \textbf{Results:}
Table \ref{tab:rq3_taxonomy} presents our derived taxonomy of patch difference patterns. We elaborate on each pattern as follows:

\ding{182} \textit{No Alignment (20.8\%).} This pattern refers to the cases where there is no alignment of automatic semantic changes between the suspicious and oracle patches.
Figure \ref{fig:rq3_example_very_different} illustrates an example.
In this example, the suspicious patch changes the base class of \verb|IsNull| and adds a new class member, 
while the oracle patch modifies the condition under which an error should be appended to the \verb|errors| list. 
The two changes are not aligned.

\ding{183} \textit{Supplementary Sem-Change (27.3\%):} This pattern refers to the cases where an unaligned Sem-change in the suspicious patch directly leads to the observed behavioral discrepancies.
In other words, the suspicious patch alters the program behavior in a way that the oracle patch does not.
Figure \ref{fig:rq3_supplementary} illustrates an example, where the change of handling input 0 in \verb|_check_vector| does not align with any change in the oracle patch. 
This category is further divided into two subcategories: 
\emph{Explicitly Handling More Possible Situations (16.9\%)}, where a behavior added only in the generated patch leads to the observed behavioral discrepancies (e.g., explicitly handling inputs of special cases, extra safety checks); 
and \emph{Supplementary Change of Application Logics (10.4\%)}, where a behavior modified only in the generated patch leads to the observed behavioral discrepancies (e.g., modifying the algorithm to calculate a value, which stays unchanged under the oracle patch).

\ding{184} \textit{Absent Sem-Change (5.2\%).} This is the opposite of the second category, where an unaligned sem-change in the oracle patch directly leads to the observed behavioral discrepancies.
Figure \ref{fig:rq3_absent} presents an example.
In this example, the sem-changes of preventing deprecation warning when \verb|values| is empty in function \verb|convert| are aligned,
while an unaligned change in oracle patch which modifies another function \verb|update| is not found in the suspicious patch. 

\ding{185} \textit{Divergent Implementations of Sem-Change (46.8\%).} In this case, different implementations of aligned sem-changes directly lead to the observed behavioral discrepancies. 
Figure \ref{fig:rq3_example_atomic} illustrates an example.
In this case, the left sem-change only adds the comma when the \verb|arg| is not empty, while the right one always adds the comma. 
Note that the upper part of the oracle patch is for refactoring, and thus is not a sem-change. 

We can see from Table~\ref{tab:rq3_taxonomy} that behavioral differences between suspicious and oracle patches are often due to \emph{Divergent Implementations of Sem-Change} (46.8\%) and due to \emph{Supplementary Sem-Change} (27.3\%).
Interestingly, there are much more suspicious patches in \emph{Supplementary Sem-Change} than in \emph{Absent Sem-Change} (27.3\% v.s. 5.2\%).
This suggests that suspicious patches tend to introduce additional changes rather than omitting necessary changes. 
In addition, 
16.9\% of the suspicious patches in \emph{Supplementary Sem-Change} add extra behaviors to explicitly handle more possible situations. 
Although these additional behaviors could make the suspicious patch more robust, they could also be unnecessary, violate user requirements, or even introduce new defects.

\vskip 1mm
\noindent \fbox{
	\parbox{0.95\linewidth}{\textbf{Answers to RQ3:} Behavioral differences between suspicious and oracle patches are often due to similar, but divergent implementations (46.8\%) and due to suspicious patches contain more semantic changes than the oracle patches. There are much more cases where the suspicious patches introduce supplementary semantic changes than those with absent semantic changes (27.3\% vs. 5.2\%). }
}

\input{tex/rq4}

%% file: tex/rq1.tex
\subsection{RQ1: Impact of Executing All Developer Tests}
As described in Section~\ref{sec:background}, the validation process of SWE-bench (Verified) assesses patch correctness only based on modified test files, which can lead to plausible but incorrect patches.
However, the severity of this flaw remains unclear. 
In this RQ, we aim to systematically evaluate the impact of this flaw on the reported performance of issue-solving tools.

\noindent \textbf{Approach:}
To answer RQ1, for each generated plausible patch,  we first apply the test patch to the repository and collect all available test files from the repository.
Next, we run these test files sequentially after applying the generated patch or the oracle patch separately. 
We then compare the test results of the two patches and record any generated patch that shows inconsistent behavior with the corresponding oracle patch, i.e., at least one test passes with the oracle patch but fails with the generated patch.
We further execute the tests that trigger inconsistent behaviors 20 times with the oracle patch to filter out flaky tests.
We observe that some developer tests focus on coding conventions rather than functionality, e.g., detecting trailing whitespace or ensuring files end with no more than one newline.
To prioritize functional correctness, we exclude the generated patches that only show inconsistencies related to coding conventions. 
The remaining generated patches introduce regression errors and are therefore incorrect.

\input{tex/tables/rq1}

\noindent \textbf{Results:}
Executing all available developer tests exposes notable overestimation in the reported performance of issue-solving tools. 
Among the plausible patches generated by the three tools, 7.2\% to 8.4\% of them are functionally incorrect when subjected to all developer tests. 
This translates to an absolute drop of 3.8\% to 5.2\% in reported resolution rates, highlighting the systematic overestimation inherent in the current validation process. 
These findings confirm the existence of plausible but incorrect patches on SWE-bench Verified and underscore the necessity of leveraging all available developer tests to enable a more robust and comprehensive assessment of patch correctness.

An example is the plausible patch produced by CodeStory to resolve django-13279. 
This issue is to use legacy \verb|encode| function to decode session data when \verb|DEFAULT_HASHING_ALGORITHM| is set to \verb|sha1|. 
Although the generated patch passes the tests in the PR-modified test files, 
it fails on another developer test where the \verb|_legacy_decode| function is tested, 
suggesting that the implementation of legacy \verb|encode| in the generated patch is not compatible with the original \verb|_legacy_decode| function. 

\noindent \fbox{
	\parbox{0.95\linewidth}{\textbf{Answers to RQ1}: Executing all developer tests reveals that on average 7.8\% of plausible patches are incorrect, which leads to an absolute performance drop of 4.5\% on average. This emphasizes the necessity of leveraging all developer tests for more robust patch validation. }
}

%% file: tex/tables/rq1.tex
\begin{table}[htbp]
    \caption{Incorrect patches detected by running all developer tests}
    \vspace{-0.25cm}
    \label{tab:rq1}
    \scalebox{0.88}{
    \begin{tabular}{cccc}
      \toprule
      \textbf{Tool} & \textbf{\%Resolved} & \textbf{\#Incorrect} & \textbf{Updated \%Resolved}\\
    \midrule
    CodeStory & 62.2\% (311/500) & 26 (8.4\%) & 57.0\% ($\downarrow$5.2\%)\\
    LearnByInteract & 60.2\% (301/500) & 23 (7.6\%) & 55.6\% ($\downarrow$4.6\%)\\
    OpenHands & 53.0\% (265/500) & 19 (7.2\%) & 49.2\% ($\downarrow$3.8\%)\\
      \bottomrule
    \end{tabular}}
\end{table}

%% file: tex/rq2.tex
\subsection{RQ2: Revealing Behavioral Discrepancies Between Plausible and Oracle Patches}~\label{sec:rq2}

\vspace{-0.2cm}

\subsubsection{Differential Patch Testing with \appname.}
The findings in RQ1 indicate that the test suites used in SWE-bench's validation process is weak.
Although the flaw mentioned in RQ1 is easy to fix,
it remains unclear whether using all developer tests is good enough to avoid plausible but incorrect patches.
To investigate this question, our insight is that if a plausible patch is incorrect, it must behave differently from its oracle patch in some scenarios.
Thus exposing and analyzing behavioral discrepancies between plausible patches and their corresponding oracle patches can facilitate the identification and understanding of plausible but incorrect patches.

\noindent \textbf{Approach:}
We employ \appname to generate tests to reveal behavioral discrepancies between each generated plausible patch and its corresponding oracle patch.
As discussed in Section~\ref{subsubsec:target-func}, we focus on the tests covering target functions. 
We refer to the tests generated by \appname as \emph{differentiating tests}, and the plausible patches with differentiating tests as \emph{suspicious patches}.

\vspace{-5pt}

\input{tex/tables/rq2_count}

\noindent \textbf{Results:}
Table \ref{tab:rq2_1} presents the number of suspicious patches generated by each evaluated tool.
Based on the tests generated by \appname, on average 29.6\% of the plausible patches are identified as suspicious patches, 
indicating that a substantial portion of plausible patches exhibit behavioral discrepancies from their oracle patches. 
If we filter out suspicious patches, the resolution rates of the three tools drop by 17.3\%, on average.
Notably, although the resolution rate of LearnByInteract (60.2\%) is higher than that of OpenHands (53.0\%), it also generates more suspicious patches than OpenHands (97 vs.\ 72), significantly reducing their difference in resolution rates (from 7.2\% to 2.2\%).
This further raises concerns about the robustness of SWE-bench.
Moreover, among the suspicious patches generated by each tool, on average 82.7\% of them cannot be identified by running all developer tests, indicating that the generated differentiating tests complement developer tests regarding identifying suspicious patches.

\input{tex/tables/rq2_cost}

Table \ref{tab:rq2_cost} details the API costs of \appname. 
To get more insightful results in the empirical study, \appname repairs the failed tests for up to 2 times, leading to a cost of 0.105\$ per patch. 
A lower repair iteration bound can reduce the cost to a minimum of 0.039\$ per patch, while only introducing an acceptable decrease (3.6\%) of suspicious patches. 
These results indicate that \appname is both effective and cost-efficient for large-scale use.

\input{tex/tables/rq2_llms}

\vspace{-5pt}

\subsubsection{Evaluating \appname with Alternative LLMs.}
To further demonstrate that \appname is not tied to a specific underlying LLM, 
we also evaluate its performance with two open-source models, \texttt{deepseek-v3} and \texttt{qwen3-235b-a22b-instruct-2507}. 
For this purpose, we randomly sample 100 plausible patches from each of the three evaluated tools and apply \appname using these models. 

\noindent \textbf{Results:}Table~\ref{tab:rq2_llms} presents the results. 
When powered by more capable LLMs, \appname achieves substantially higher detection rates of suspicious patches, 
showing that its effectiveness generalizes across different underlying models. 
This confirms that \appname is not coupled with GPT-4o-mini and can work with open-source models.
Nevertheless, for the main study, we adopt GPT-4o-mini as the underlying model because it offers both low cost and high usability.

\subsubsection{Comparison with Existing Test Generators}~\label{subsec:comparison}
The tests generated by existing test generators may also reveal behavioral differences between patches.
However, these generators aim to generate regression tests to cover as much of the code under test as possible, while our approach targets patch-modified code and focuses on revealing behavioral discrepancies between patches.
We compare \appname with existing test generators to evaluate its effectiveness.

\noindent\textbf{Approach:} We use two representative test generators for Python (i.e., Pynguin~\cite{lukasczyk2022pynguin} and CoverUp~\cite{pizzorno2024coverup}), 
along with an issue reproduction tool LIBRO~\cite{kang2024evaluating}.
Pynguin is the state-of-the-art search-based test generation tool for Python.
CoverUp is a representative LLM-based test generation tool with its implementation publicly available and easy to use.
Please note that both Pynguin and CoverUp support only Python 3.10 or higher, a requirement met by only 14.8\% of the 500 tasks in SWE-bench Verified.
Consequently, our evaluation is confined to the plausible patches generated by the three tools that are under a compatible Python version, yielding 133 plausible patches.
For both test generators, we designate the patch-modified files as the target modules, and tests are generated on the repository with the oracle patch applied. 
Since CoverUp supports only PyTest and fails on the instances employing other test frameworks (e.g., Django’s customized framework), we do not provide original tests to CoverUp to ensure it can handle the 133 instances.
Since LIBRO was originally implemented in Java, we re-implemented it in Python. 
We omit its selection phase, as it only ranks tests for developers and does not potentially increase differentiating tests; 
instead, we execute all generated tests to check for behavioral discrepancies between each plausible patch and its oracle. 
We use the best-performing configuration from the original paper (two examples, $n=50$)~\cite{kang2024evaluating}. 
For both CoverUp and LIBRO, we follow \appname’s setting and adopt \texttt{gpt-4o-mini-2024-07-18} as the underlying LLM for fair comparison.

\input{tex/tables/baseline_1}

\noindent \textbf{Results:}
Table ~\ref{tab:baseline_test_gen} presents the evaluation results for Pynguin and CoverUp.
Pynguin produces at least one test file for only 3 plausible patches 
because it frequently encounters exceptions during generation.
For example, for 93 patches, Pynguin raises an \verb|AttributeError| which states that in the file \verb|pynguin/testcase/execution.py|, attribute \verb|IN| cannot be found in enum \verb|Compare|, which may be a bug in Pynguin. %
CoverUp produces at least one test file only for 40 patches due to the difficulties it has in constructing appropriate execution environments (such as the database setup required by Django).
None of the tests generated by Pynguin or CoverUp reveal behavioral discrepancies between the plausible and oracle patches, i.e., none of these tests are differentiating tests.
In contrast, \appname produces at least one test file for 117 patches and successfully generates differentiating tests for 56 of them.
Note that among the 133 plausible patches, \appname only attempts to generate tests for 117 of them.
15 of them are identical to their corresponding oracle patches and are omitted by \appname, and 1 patch contains syntax errors and cannot be correctly parsed by the unidiff library~\cite{unidiff}.
Table~\ref{tab:rq2_libro} summarizes the evaluation results for LIBRO.
Even under its best-performing configuration, LIBRO is able to generate differentiating tests for only 5.9\% of the plausible patches, 
in contrast to 29.6\% achieved by \appname. 
This limited performance can be attributed to LIBRO’s reliance solely on issue descriptions for test generation, 
without considering the actual patch content. 
By contrast, \appname explicitly targets the patch-relevant code and aims to generate tests that expose subtle behavioral differences between patches. 
These results underscore the effectiveness and necessity of \appname in generating differentiating tests.

\input{tex/tables/rq2_libro}

\vskip 1mm
\noindent \fbox{
	\parbox{0.95\linewidth}{\textbf{Answers to RQ2}: \appname generates differentiating tests for 29.6\% of the plausible patches. This suggests that a significant proportion of plausible patches behave differently from the oracle patches. In contrast, test generation baselines can only generate differentiating tests for up to 5.9\% of the plausible pathces.}
}

%% file: tex/tables/rq2_count.tex
\begin{table}[htbp]
\centering
\caption{The number and the impact of the suspicious patches generated by CodeStory, LearnByInteract, and OpenHands.}
\vspace{-0.4cm}
\scalebox{0.92}{
\begin{threeparttable}
\begin{tabular}{cccccl}
\toprule
\multirow{2}{*}{\textbf{Tool}} & \multirow{2}{*}{\textbf{\%Resolved}} & \multicolumn{2}{c}{\textbf{\#Patches}} & \textbf{\%Resolved} \\
& & \textbf{Susp.} & \textbf{Uniq.} & \textbf{w/o Susp.} \\
\midrule
CodeStory & 62.2\% (311) & 91 (29.3\%) & 74/91 & 44.0\% ($\downarrow$18.2\%) \\
LearnByInteract & 60.2\% (301) & 97 (32.2\%) & 81/97 & 40.8\% ($\downarrow$19.4\%) \\
OpenHands & 53.0\% (265) & 72 (27.2\%) & 60/72 & 38.6\% ($\downarrow$14.4\%) \\
\bottomrule
\end{tabular}
\begin{tablenotes}
\footnotesize
\item \noindent\textbf{Susp.} refers to suspicious. \textbf{Uniq.} patches refer to the suspicious patches not identified by running all developer tests.
\end{tablenotes}
\end{threeparttable}
}
\vspace{-0.2cm}
\label{tab:rq2_1}%
\end{table}

%% file: tex/tables/rq2_cost.tex
\begin{table}[htbp]
    \caption{API costs of \appname under different repair iteration bounds}
    \vspace{-0.25cm}
    \label{tab:rq2_cost}
    \scalebox{0.88}{
    \begin{tabular}{cccc}
      \toprule
      \textbf{Max Repair Iter} & \textbf{\#Susp. Patches} & \textbf{Cost/Patch(\$)} & \textbf{Total Cost (\$)}\\
    \midrule
    2 & 260/877 (29.6\%) & 0.105 & 91.716\\
    1 & 250/877 (28.5\%) & 0.075 & 66.186\\
    0 & 228/877 (26.0\%) & 0.039 & 33.962\\
      \bottomrule
    \end{tabular}}
\end{table}

%% file: tex/tables/rq2_llms.tex
\begin{table}[htbp]
    \caption{Detected suspicious patches using different underlying LLMs (sampled 100 patches for each tool)}
    \vspace{-0.25cm}
    \label{tab:rq2_llms}
    \scalebox{0.892}{
    \setlength{\tabcolsep}{3pt}
    \begin{tabular}{@{}ccccc@{}}
      \toprule
      \textbf{Model} & \textbf{CodeStory} & \textbf{LearnByInteract} & \textbf{OpenHands} & \textbf{Overall}\\
    \midrule
    GPT-4o-mini & 26 (26.0\%) & 32 (32.0\%) & 26 (26.0\%) & 84 (28.0\%)\\
    DeepSeek-V3 & 34 (34.0\%) & 43 (43.0\%) & 40 (40.0\%) & 117 (39.0\%)\\
    Qwen3-A22B & 45 (45.0\%) & 49 (49.0\%) & 49 (49.0\%) & 143 (47.7\%)\\
      \bottomrule
    \end{tabular}}
    \vspace{-0.3 cm}
\end{table}

%% file: tex/tables/baseline_1.tex
\begin{table}[htbp]
  \caption{Comparing \appname with test generation tools. Gen. Tests and Diff. Tests refer to generated tests and differentiating tests, respectively.}
  \label{tab:baseline_test_gen}

  \vspace{-0.3cm}
  \scalebox{0.9}{
  \begin{tabular}{ccc}
    \toprule
    \textbf{Tool} & \textbf{\#Patches /w Gen. Tests} & \textbf{\#Patches /w Diff. Tests} \\
    \midrule
    \appname & \textbf{117} / 133 & \textbf{56} / 133 \\
      Pynguin   & 3 / 133  & 0 / 133 \\
      CoverUp   & 40 / 133 & 0 / 133\\
    \bottomrule
  \end{tabular}}
\end{table}

%% file: tex/tables/rq2_libro.tex
\begin{table}[htbp]
    \caption{Suspicious patches detected by \appname and LIBRO}
    \vspace{-0.25cm}
    \label{tab:rq2_libro}
    \scalebox{0.892}{
    \setlength{\tabcolsep}{3pt}
    \begin{tabular}{ccccc}
      \toprule
      \textbf{Tool} & \textbf{CodeStory} & \textbf{LearnByInteract} & \textbf{OpenHands} & \textbf{Overall}\\
    \midrule
    \appname & 91 (29.3\%) & 97 (32.2\%) & 72 (27.2\%) & 260 (29.6\%)\\
    LIBRO & 18 (5.8\%) & 20 (6.6\%) & 14 (5.3\%) & 52 (5.9\%)\\
      \bottomrule
    \end{tabular}}
\end{table}

%% file: tex/tables/rq3_taxonomy.tex
\begin{table}[t]
\centering
\caption{Taxonomy for patterns of patch differences that lead to behavioral discrepancies. Sem-change refers to atomic semantic change. }
\label{tab:rq3_taxonomy}

\vspace{-0.25cm}

\setlength{\defaultaddspace}{3pt}  %
\scalebox{0.95}{
\begin{tabular}{@{}>{\RaggedRight}p{7cm}l@{}}
\toprule
\textbf{Category} & \textbf{\#Cases} \\
\midrule
\textbf{Total}                                                  & \textbf{77}\\
\hdashline
\addlinespace[4pt]

\textbf{Absent Sem-Change}       & \textbf{4 (5.2\%)} \\
\addlinespace[4pt]

\textbf{Supplementary Sem-Change} & \textbf{21 (27.3\%)} \\
\hspace{3mm}\textit{-- Explicitly Handling More Possible Situations} & 13 (16.9\%) \\
\hspace{3mm}\textit{-- Supplementary Change of Application Logics}     & 8 (10.4\%) \\
\addlinespace[4pt]

\textbf{Divergent Implementations of Sem-Change} & \textbf{36 (46.8\%)}\\
\addlinespace[4pt]

\textbf{No Alignment}      & \textbf{16 (20.8\%)} \\
\bottomrule
\end{tabular}}
\end{table}

%% file: tex/rq4.tex
\subsection{RQ4: The Correctness of Suspicious Patches}

Suspicious patches are not necessarily incorrect patches.
For example, the program behavior with illegitimate values as inputs can be undefined.
If a differentiating test triggers different behaviors with such illegitimate inputs, the corresponding suspicious patch can also be correct.
In this RQ, we aim to analyze the correctness of suspicious patches and identify the reasons for correct and incorrect suspicious patches.

\noindent \textbf{Approach:}
We utilize the suspicious patches sampled in RQ3.
For each sampled suspicious patch, the first author (A1) manually analyzes the user requirements in the issue statement, the results of running all developer-written regression tests (identified in RQ1), and the results of running the generated differentiating tests, and then compares the implementations of the suspicious and oracle patches.
A patch is considered correct if it satisfies the requirements specified in the issue statement without introducing errors.
In cases of ambiguity, A1 discusses with another author (A2) until a consensus is reached.

\input{tex/tables/rq4_correctness}

\noindent \textbf{Results:}
After manual inspection, each sampled patch is classified as incorrect, correct, or with uncertain correctness, as summarized in Table ~\ref{tab:rq4_correctness}.

\textbf{Incorrect Patches (22):} We identify 28.6\% of the suspicious patches as incorrect, which can further be divided into four sub-categories.
\ding{182} \textit{Regressive Patches (11):} 
These patches successfully satisfy the requirements mentioned in the issue statements, but accidentally introduce faults into some functionalities that are unrelated to the target issues. 
Our motivating example in Subsection \ref{subsec:example} falls into this category, where the issue is to fix the bug that the \verb|ValueError| of "Imaginary coordinates are not permitted" is raised under \verb|with evaluate(False)| when there is no imaginary input. 
The generated patch satisfies the user requirement that this \verb|ValueError| is now not raised under ordinary inputs. 
However, it also makes the object creation never raise this \verb|ValueError| when \verb|evaluate| is set to \verb|False|, 
even though there is an imaginary input. 
This violates the original functionality, therefore this patch falls into Regressive Patches. 
\ding{183} \textit{Partial Fixes (6):} 
These patches partially resolve the target issues but fail to satisfy the requirements in the issue statements for certain legitimate inputs.
For example, the patch produced by LearnByInteract to resolve scikit-learn-14496 aims to fix a bug where the \verb|OPTICS::fit| function raises \verb|TypeError| when the \verb|min_samples| attribute of \verb|OPTICS| is set to a float between 0 and 1. 
However, the patch fails to locate and fix every piece of relevant buggy code and makes \verb|OPTICS::fit| raise \verb|ValueError| under certain \verb|min_samples| between 0 and 1, violating the issue statement.
\ding{184} \textit{Patches with Irrelevant Behavioral Changes (3):} 
These patches successfully satisfy the requirements mentioned in the issue statements but also introduce some unnecessary changes. 
For example, the patch produced by LearnByInteract for django-15104 changes a code line to safely remove the key \verb|to| from a dictionary when it exists, which successfully fixes the bug. 
However, this patch additionally repeats another code line \verb|fields_def.append(deconstruction)|, 
which introduces duplicate items in \verb|fields_def|. 
Different from \emph{Regressive Patches}, which introduces regression in their implementation to resolve the issue, 
patches in this category contain changes that do not contribute to resolving the issue. 
\ding{185} \textit{Patches with Erroneous Modifications (2):} 
These patches contain clear implementation errors. 
For example, the patch produced by CodeStory to resolve sympy-20801 captures exceptions of type \verb|SympifyError|, which is not defined before. 
These patches are plausible because the original test suites fail to cover the buggy branches and do not trigger runtime errors.

By sampling 30\% of the suspicious patches for manual validation, we identify 22 incorrect patches.
If we assume that the incorrect patches are distributed evenly among suspicious patches and consider the incorrect patches that are discovered in RQ1 but are not suspicious (which results in 23 patches), the estimated incorrect rate will be 11.0\% among plausible patches. 
This leads to the resolution rates of the studied tools being inflated by 6.4 points on average, 
further confirming the prevalence of performance overestimation and necessitating tools like \appname for detecting plausible but incorrect patches. 
In addition, 11 (50.0\%) of the 22 incorrect patches cannot be identified by running all developer tests, underscoring \appname's effectiveness in helping identify incorrect patches.
Among the 22 incorrect patches, 8 (36.4\%) patches are attributed to faulty supplementary semantic changes.
This highlights the risk of supplementary changes in plausible patches and emphasizes the need for issue-solving tools to rigorously review and refine the plausible patches.

\textbf{Correct Patches (4):} Only 5.2\% patches are certainly correct, which can be divided into two sub-categories.
\ding{182} \textit{Irrelevant Oracle Changes (2):}
In these cases, the observed behavioral discrepancy arises from the changes in oracle patches that do not contribute to issue resolution.  
So there is no evidence suggesting the plausible patch fails to address the issue. 
\ding{183} \textit{Invalid Differentiating Tests (2):}
In these cases, the differentiating tests rely on illegitimate inputs where the program's expected behavior is undefined.
Such tests should not lead to incorrect patches. 

\textbf{Patches with Uncertain Correctness (51):}
For 66.2\% of the sampled suspicious patches, we cannot determine their correctness based on the information that we can find from SWE-bench and the corresponding repositories.
This uncertainty arises due to the differences in implementation details between the plausible and oracle patches that are not explicitly specified in the issue statement. 
In such cases, neither the issue-solving tools nor our analysis can reliably infer user requirements regarding these implementation details. 
The plausible patch generated by CodeStory to solve matplotlib-25311 falls in this category.  
The issue aims at fixing a bug where pickling figures raises "TypeError: cannot pickle \verb|FigureCanvasQTAgg| object". 
This plausible patch introduces \verb|__getstate__| and \verb|__setstate__| functions to ensure that the unpickable attribute \verb|canvas| is set to \verb|None| during serialization and remains \verb|None| upon deserialization, which is a common practice for handling such issues. 
The oracle patch, however, transforms \verb|canvas| into a property using a lambda expression, so that the value of \verb|canvas| is dynamically retrieved when accessed and not stored during pickling. 
The key difference is that the plausible patch explicitly sets \verb|canvas| to \verb|None| upon loading, whereas the oracle patch makes it remain accessible after deserialization. 
Since there are no explicit requirements dictating whether \verb|canvas| should remain accessible, we cannot determine the correctness of this plausible patch. 
While some unspecified details may have negligible or no impact, others could lead to unintended behaviors or latent issues.
These findings underscore a need to enhance issue-solving tools with the capabilities to detect ambiguous or under-specified requirements and to proactively prompt users for clarification.
They also imply that the community may need a better benchmark, where issues are well specified and leave less ambiguity.

\vskip 1mm
\noindent \fbox{
	\parbox{0.95\linewidth}{\textbf{Answers to RQ4:} Among the manually validated 77 suspicious patches, 22 (28.6\%) patches are incorrect. This interprets to an estimated incorrect rate of 11.0\% in plausible patches, inflating the reported resolution rate by 6.4 points on average. 51 (66.2\%) patches have uncertain correctness due to under-specified requirements.}
}

%% file: tex/tables/rq4_correctness.tex
\begin{table}[t]
\centering
\caption{Results of manually validating patch correctness }
\label{tab:rq4_correctness}
\vspace{-0.3cm}

\setlength{\defaultaddspace}{3pt}
\scalebox{0.93}{
\begin{tabular}{@{}>{\RaggedRight}p{7.4cm}l@{}}
\toprule
\textbf{Category} & \textbf{\#Cases} \\
\midrule

\textbf{Total}                                                  & \textbf{77}\\
\textbf{Incorrect Patches Detected in RQ1}                      & \textbf{11 (14.3\%)}\\
\hdashline
\addlinespace[4pt]

\textbf{Incorrect Patches}                                        & \textbf{22 (28.6\%)}\\
\hspace{3mm}\textit{-- Regressive Patches}                        & 11 (14.3\%)\\
\hspace{3mm}\textit{-- Partial Fixes}                             & 6 (7.8\%)\\
\hspace{3mm}\textit{-- Patches with Irrelevant Behavioral Changes}  & 3 (3.9\%)\\
\hspace{3mm}\textit{-- Patches with Erroneous Modifications}       & 2 (2.6\%)\\
\addlinespace[4pt]

\textbf{Correct Patches}                                  & \textbf{4 (5.2\%)}\\
\hspace{3mm}\textit{-- Irrelevant Changes in Oracle Patch}           & 2 (2.6\%)\\
\hspace{3mm}\textit{-- Invalid Differentiating Tests}   & 2 (2.6\%)\\
\addlinespace[4pt]

\textbf{Patches with Uncertain Correctness}                                             & \textbf{51 (66.2\%)}\\
\addlinespace[4pt]

\bottomrule
\end{tabular}}

\smallskip

\end{table}

%% file: tex/discussion.tex
\section{Discussion}

\subsection{Implications}
\textit{Carefully selecting developer tests for robust patch validation.} As shown in RQ1, ignoring the test files not modified in the PR leads to notable performance inflation.
Yet, we also find some projects contain some non-functional tests, e.g., tests related to code conventions.
These findings indicate that maintainers of issue-solving benchmarks should carefully select developer tests for patch validation.
One suggestion is to use all developer tests by default and exclude non-functional tests for benchmarks only focusing on functional correctness.

\textit{Awareness of plausible but incorrect patches.} The results on RQ4 show that the patches that pass all developer tests can still be incorrect. This suggests that both the users and the maintainers of SWE-bench should check plausible patches generated by issue-solving tools and filter out incorrect patches for more accurate evaluation, as is common in automated program repair~\cite{cacm2019-program-repair}.

\textit{Paying more attention to patches introducing supplementary semantic changes.}
As shown in RQ4, 36.4\% of the certainly incorrect patches are attributed to faulty supplementary semantic changes, suggesting that additional changes can be an indicator for incorrectness. Thus, the users and maintainers of SWE-bench should pay more attention to patches with supplementary semantic changes when checking the correctness of patches.

\textit{Handling under-specified issue statements.}
The result of RQ4 shows that a large proportion of suspicious patches have uncertain correctness due to under-specified requirements in the issue statement.
These patches potentially introduce unintended or undesired behaviors, compromising the robustness of the target project. 
This calls for better issue-solving tools that are capable of detecting and refining under-specified requirements collaboratively with users.

\textit{Building a new benchmark with well-specified statements.}
Although the issue statements in SWE-bench are considered to be of high quality by human annotators~\cite{swebenchverified}, vague issue statements with under-specified requirements still exist, as shown in RQ4.
Such issue statements can misguide issue-solving tools in generating incorrect patches, constraining the reliability of the benchmark. 
Practitioners should consider building better benchmarks, where issues are well-specified and leave less ambiguity.

\subsection{Towards Sustainable Patch Validation in SWE-bench}

We envision \appname being useful for sustainably strengthening SWE-bench and other issue-solving benchmarks.
For users of SWE-bench, 
before submitting the patches produced by their tools to the SWE-bench leaderboard, 
it is advisable to assess the plausible patches by utilizing \appname to generate differentiating tests and manually examine if the behavioral discrepancies exposed by tests lead to incorrect patches.
This would lead to a more accurate evaluation.
While such a practice undoubtedly imposes additional effort and cost on users, 
it brings significant long-term benefits. 
Once a user identifies plausible but incorrect patches with \appname, 
she can submit the differentiating tests that expose incorrectness to the SWE-bench leaderboard. 
Incorporating these tests into the benchmarks's original test suite enables the identification of similar incorrect patches in the future, thus strengthening the validation process.
This collaborative contribution facilitates a more rigorous validation process for subsequent users.
Over time, as the test suite continues to evolve and becomes comprehensive, the additional burden would also decrease. 
This fosters a sustainable and robust patch validation ecosystem for assessing issue-solving tools.

\subsection{Threats to Validity}

Our study is subject to several potential threats to validity and intrinsic limitations: 
1) To balance costs and effectiveness, the implementation of \appname leverages GPT-4o-mini as the underlying LLM. 
While this choice limits access to potentially more advanced models, 
\appname successfully generates differentiating tests for 29.6\% of the plausible patches, 
establishing a solid basis for later studies. 
2) The analysis conducted in RQ3 and RQ4 involves manual analysis, potentially introducing human bias. 
To reduce this risk, the first author collaborates with another author to resolve ambiguous cases and reach a consensus on uncertain assessments, thereby enhancing objectivity.
3) For RQ3 and RQ4, we sample 30\% (77) of the suspicious plausible patches for analysis.
This sample size may introduce sampling bias.
However, this sample size corresponds to a confidence interval of 9.39\% with a confidence level of 95\%.
Prior work~\cite{linares2014api,hassan2018studying} shows that it is sufficiently large to draw statistically meaningful conclusions.
4) RQ3 and RQ4 are restricted to the suspicious patches identified by \appname, potentially introducing bias as this subset may not fully represent the characteristics of all plausible patches.
Despite threats 3) and 4), 
the observed trends and patterns still provide meaningful evidence regarding the severity of performance overestimation and offer actionable insights into patch correctness validation. 

\vspace{-0.1cm}

%% file: tex/related_work.tex
\section{Related Work}

\paragraph{Weak test suites}
The problem of weak test suites has been well studied in the area of test-based automatic program repair (APR)~\cite{martinez2015automatic, qi2015analysis, smith2015cure, liu2021critical}.
For instance, 
Martinez et al.~\cite{martinez2015automatic} manually assessed 84 plausible patches generated on the Defects4J benchmark~\cite{just2014defects4j}, and found that only 11 of them are correct. 
Qi et al. ~\cite{qi2015analysis} found that most of the plausible patches 
are equivalent to a single modification that deletes the corresponding functionality, which is problematic. 
Some prior work proposed actionable strategies to mitigate this problem in APR~\cite{smith2015cure, xiong2018identifying, yu2019alleviating, yang2023large}. 
For example, 
Smith el al.~\cite{smith2015cure} found that using a test suite that is inaccessible to evaluated tools can help detect plausible but incorrect patches.
Xiong et al.~\cite{xiong2018identifying} proposed to incorporate execution behavior similarity under generated test inputs to detect incorrect patches that behave significantly differently from their oracle patches.
For test-based APR benchmarks, test suites are provided to the evaluated tools, so plausible but incorrect patches are also called overfitting patches.
Different from these studies, our work focuses on the issue-solving task, which aims to resolve issues based on issue statements rather than test suites.
For issue-solving benchmarks, the evaluated tools cannot access test suites during evaluation.
Thus, prior conclusions may not apply to SWE-bench.
Recently, Aleithan et al.~\cite{aleithan2024swe} also noticed the weak test suite problem in SWE-bench.
However, first, their study is conducted on the patches generated by only one issue-solving tool on SWE-bench full, which is shown to contain many low-quality instances.
In contrast, our study covers three state-of-the-art tools and focuses on the human-filtered subset SWE-bench Verified.
Second, they focus on manually classifying plausible patches, while we dig deeper into the patterns and types of plausible but incorrect patches with our novel technique \appname.

\paragraph{Automated test generation}
Various automated test generation approaches are proposed to alleviate testing efforts and help find bugs\cite{fraser2011evosuite, lukasczyk2022pynguin, pacheco2007randoop, agitar2023}.
\emph{Traditional test generation} approaches utilize some predefined rules to generate tests and can be mainly categorized into search-based~\cite{fraser2011evosuite, lukasczyk2022pynguin}, randomization-based~\cite{pacheco2007randoop}, and constraints-based~\cite{tillmann2008pex, chang2016constraint} approaches.
\emph{Deep learning-based test generation} approaches train deep learning models to generate natural and human-understandable test cases~\cite{tufano2020unit, alagarsamy2024a3test, dinella2022toga, rao2023cat}.
Recently, researchers have proposed \emph{LLM-based test generation} approaches~\cite{chen2024chatunitest, ryan2024code, pizzorno2024coverup, wang2024hits}.
For example, CoverUp iteratively instructs LLMs to generate Python regression tests, improve coverage, and fix errors with detailed coverage information.
Liu et al. ~\cite{liu2024your} proposed to augment the test suites in the code generation benchmarks HumanEval~\cite{chen2021evaluating} and MBPP~\cite{austin2021program} based on LLMs and type-aware mutation.
These approaches aim to generate regression tests to cover as much of the code as possible, while \appname targets differentiating two patches, which requires more targeted test generation.
Differential testing aims at generating tests to expose different behaviors between two versions of a program ~\cite{evans2007differential, li2023nuances, liu2024llm, etemadi2024mokav}. 
Mokav~\cite{etemadi2024mokav} uses execution feedback to iteratively guide an LLM in generating tests to differentiate different solutions of programming competition tasks.
Existing LLM-based differential testing techniques focus on function-level programs, while \appname can generate differentiating tests for large and complex projects.
In addition, \appname targets differential patch testing, and leverages specially designed methods to identify appropriate target functions and to construct useful contextual information for revealing meaningful behavioral differences, which are important for differential patch testing but are not considered by prior work.

\paragraph{Software engineering agents}
LLM-based agents promise to help automate various software engineering tasks~\cite{AgenticAISE2025}.
Agents for various tasks have been presented, such as fault localization~\cite{kang2024evaluating,xu2025flexfl,jiang2025cosil}, issue solving~\cite{Yang2024a,zhang2024autocoderover,xia2024agentless}, automated program repair~\cite{icse2025-RepairAgent,Cheng2025}, repository setup~\cite{issta2025_ExecutionAgent,Eliseeva2025,Milliken2025}, and generating issue-reproducing tests~\cite{Muendler2024,Issue2Test_arXiv2025}.
Researchers have started to study the behavior such agents to better understand their strengths and weaknesses~\cite{ase2025_agent_study}.
Our work complements such efforts and will help improve future software engineering agents by critically analyzing the patches produced by the agents.

%% file: tex/conclusion.tex
\section{Conclusion}
This paper presents an in-depth empirical study of the correctness of plausible generated patches on SWE-bench.
We first find that the validation process of SWE-bench overlooks non-modified test files, which leads to significant performance overestimation. 
Then we extensively investigate the prevalence of plausible patches that exhibit behavioral discrepancies from ground truth patches, the patch difference patterns leading to behavioral discrepancies, and the correctness of plausible patches that exhibit behavioral discrepancies.
The core of our methodology is the novel \appname technique for differential patch testing, which automatically exposes behavioral discrepancies between two patches.
The results call for carefully selecting developer tests for patch validation, checking and filtering out plausible but incorrect patches for more accurate evaluation, and paying more attention to supplementary semantic changes in plausible patches.
Our work will contribute toward better issue-solving tools and benchmarks that address the problem of under-specified specifications.
\appname can provide concrete evidence for invalid functionality in patches through generated tests, enabling more focused and objective patch assessment.
The test generated by \appname can be continuously used to strengthen issue-solving benchmarks with the help of benchmark users.
We envision \appname to be useful for building a sustainable and robust patch validation ecosystem for assessing issue-solving tools.

%% file: tex/data.tex
\section{Data Availability}

Our code and data are available: \url{https://github.com/ZJU-CTAG/PatchDiff}

%% file: tex/acks.tex
\section{Acknowledgments}

This work was supported by the National Natural Science Foundation of China (No. 62202420), Zhejiang Provincial Natural Science Foundation of China (No. LZ25F020003), the European Research Council (ERC, grant agreements 851895 and 101155832), and the German Research Foundation within the DeMoCo project.

%% file: main.bbl

\begin{thebibliography}{63}


\ifx \showCODEN    \undefined \def \showCODEN     #1{\unskip}     \fi
\ifx \showISBNx    \undefined \def \showISBNx     #1{\unskip}     \fi
\ifx \showISBNxiii \undefined \def \showISBNxiii  #1{\unskip}     \fi
\ifx \showISSN     \undefined \def \showISSN      #1{\unskip}     \fi
\ifx \showLCCN     \undefined \def \showLCCN      #1{\unskip}     \fi
\ifx \shownote     \undefined \def \shownote      #1{#1}          \fi
\ifx \showarticletitle \undefined \def \showarticletitle #1{#1}   \fi
\ifx \showURL      \undefined \def \showURL       {\relax}        \fi
\providecommand\bibfield[2]{#2}
\providecommand\bibinfo[2]{#2}
\providecommand\natexlab[1]{#1}
\providecommand\showeprint[2][]{arXiv:#2}

\bibitem[agi(2023)]%
        {agitar2023}
 \bibinfo{year}{2023}\natexlab{}.
\newblock \bibinfo{title}{agitar}.
\newblock \bibinfo{howpublished}{\url{http://www.agitar.com/}}.
\newblock
\newblock
\shownote{[Online]. Available: http://www.agitar.com/ Accessed: 2025-09-04}.


\bibitem[cla(2024)]%
        {claude-3.5}
 \bibinfo{year}{2024}\natexlab{}.
\newblock \bibinfo{title}{Introducing computer use, a new Claude 3.5 Sonnet, and Claude 3.5 Haiku}.
\newblock
\urldef\tempurl%
\url{https://www.anthropic.com/news/3-5-models-and-computer-use}
\showURL{%
\tempurl}
\newblock
\shownote{Accessed: 2025-09-04}.


\bibitem[swe(2024)]%
        {swebenchverified}
 \bibinfo{year}{2024}\natexlab{}.
\newblock \bibinfo{title}{Introducing SWE-bench Verified}.
\newblock
\urldef\tempurl%
\url{https://openai.com/index/introducing-swe-bench-verified/}
\showURL{%
\tempurl}
\newblock
\shownote{Accessed: 2025-09-04}.


\bibitem[bla(2025)]%
        {blackboxai}
 \bibinfo{year}{2025}\natexlab{}.
\newblock \bibinfo{title}{BLACKBOX.AI}.
\newblock
\urldef\tempurl%
\url{https://www.blackbox.ai/}
\showURL{%
\tempurl}
\newblock
\shownote{Accessed: 2025-09-04}.


\bibitem[cod(2025)]%
        {codestory}
 \bibinfo{year}{2025}\natexlab{}.
\newblock \bibinfo{title}{codestoryai/aide: The open-source AI-native IDE}.
\newblock
\urldef\tempurl%
\url{https://github.com/codestoryai/aide}
\showURL{%
\tempurl}
\newblock
\shownote{Accessed: 2025-09-04}.


\bibitem[iso(2025)]%
        {isoformai}
 \bibinfo{year}{2025}\natexlab{}.
\newblock \bibinfo{title}{Isoform - Get custom integrations to close B2B deals using your 24/7 integration developer}.
\newblock
\urldef\tempurl%
\url{https://www.isoform.ai/}
\showURL{%
\tempurl}
\newblock
\shownote{Accessed: 2025-09-04}.


\bibitem[gpt(2025)]%
        {gpt-o1}
 \bibinfo{year}{2025}\natexlab{}.
\newblock \bibinfo{title}{OpenAI o1 and new tools for developers}.
\newblock
\urldef\tempurl%
\url{https://openai.com/index/o1-and-new-tools-for-developers/}
\showURL{%
\tempurl}
\newblock
\shownote{Accessed: 2025-09-04}.


\bibitem[pac(2025)]%
        {package}
 \bibinfo{year}{2025}\natexlab{}.
\newblock \bibinfo{title}{Our replication package}.
\newblock
\urldef\tempurl%
\url{https://github.com/ZJU-CTAG/PatchDiff}
\showURL{%
\tempurl}
\newblock
\shownote{Accessed: 2025-09-04}.


\bibitem[har(2025)]%
        {harness-swebench}
 \bibinfo{year}{2025}\natexlab{}.
\newblock \bibinfo{title}{SWE-bench/swebench/harness at main · swe-bench/SWE-bench}.
\newblock
\urldef\tempurl%
\url{https://github.com/swe-bench/SWE-bench/tree/main/swebench/harness}
\showURL{%
\tempurl}
\newblock
\shownote{Accessed: 2025-09-04}.


\bibitem[uni(2025)]%
        {unidiff}
 \bibinfo{year}{2025}\natexlab{}.
\newblock \bibinfo{title}{The unidiff library}.
\newblock
\urldef\tempurl%
\url{https://pypi.org/project/unidiff/}
\showURL{%
\tempurl}
\newblock
\shownote{Accessed: 2025-09-04}.


\bibitem[wan(2025)]%
        {wandbai}
 \bibinfo{year}{2025}\natexlab{}.
\newblock \bibinfo{title}{Weights \& Biases: The AI Developer Platform}.
\newblock
\urldef\tempurl%
\url{https://wandb.ai/site}
\showURL{%
\tempurl}
\newblock
\shownote{Accessed: 2025-09-04}.


\bibitem[Alagarsamy et~al\mbox{.}(2024)]%
        {alagarsamy2024a3test}
\bibfield{author}{\bibinfo{person}{Saranya Alagarsamy}, \bibinfo{person}{Chakkrit Tantithamthavorn}, {and} \bibinfo{person}{Aldeida Aleti}.} \bibinfo{year}{2024}\natexlab{}.
\newblock \showarticletitle{A3test: Assertion-augmented automated test case generation}.
\newblock \bibinfo{journal}{\emph{Information and Software Technology}}  \bibinfo{volume}{176} (\bibinfo{year}{2024}), \bibinfo{pages}{107565}.
\newblock


\bibitem[Aleithan et~al\mbox{.}(2024)]%
        {aleithan2024swe}
\bibfield{author}{\bibinfo{person}{Reem Aleithan}, \bibinfo{person}{Haoran Xue}, \bibinfo{person}{Mohammad~Mahdi Mohajer}, \bibinfo{person}{Elijah Nnorom}, \bibinfo{person}{Gias Uddin}, {and} \bibinfo{person}{Song Wang}.} \bibinfo{year}{2024}\natexlab{}.
\newblock \showarticletitle{SWE-Bench+: Enhanced Coding Benchmark for LLMs}.
\newblock \bibinfo{journal}{\emph{arXiv preprint arXiv:2410.06992}} (\bibinfo{year}{2024}).
\newblock


\bibitem[Austin et~al\mbox{.}(2021)]%
        {austin2021program}
\bibfield{author}{\bibinfo{person}{Jacob Austin}, \bibinfo{person}{Augustus Odena}, \bibinfo{person}{Maxwell Nye}, \bibinfo{person}{Maarten Bosma}, \bibinfo{person}{Henryk Michalewski}, \bibinfo{person}{David Dohan}, \bibinfo{person}{Ellen Jiang}, \bibinfo{person}{Carrie Cai}, \bibinfo{person}{Michael Terry}, \bibinfo{person}{Quoc Le}, {et~al\mbox{.}}} \bibinfo{year}{2021}\natexlab{}.
\newblock \showarticletitle{Program synthesis with large language models}.
\newblock \bibinfo{journal}{\emph{arXiv preprint arXiv:2108.07732}} (\bibinfo{year}{2021}).
\newblock


\bibitem[Bouzenia et~al\mbox{.}(2025)]%
        {icse2025-RepairAgent}
\bibfield{author}{\bibinfo{person}{Islem Bouzenia}, \bibinfo{person}{Premkumar Devanbu}, {and} \bibinfo{person}{Michael Pradel}.} \bibinfo{year}{2025}\natexlab{}.
\newblock \showarticletitle{{RepairAgent}: An Autonomous, {LLM}-Based Agent for Program Repair}. In \bibinfo{booktitle}{\emph{International Conference on Software Engineering (ICSE)}}.
\newblock


\bibitem[Bouzenia and Pradel(2025a)]%
        {ase2025_agent_study}
\bibfield{author}{\bibinfo{person}{Islem Bouzenia} {and} \bibinfo{person}{Michael Pradel}.} \bibinfo{year}{2025}\natexlab{a}.
\newblock \showarticletitle{Understanding Software Engineering Agents: A Study of Thought-Action-Result Trajectories}. In \bibinfo{booktitle}{\emph{ASE}}.
\newblock


\bibitem[Bouzenia and Pradel(2025b)]%
        {issta2025_ExecutionAgent}
\bibfield{author}{\bibinfo{person}{Islem Bouzenia} {and} \bibinfo{person}{Michael Pradel}.} \bibinfo{year}{2025}\natexlab{b}.
\newblock \showarticletitle{You name it, I run it: An LLM agent to execute tests of arbitrary projects}.
\newblock \bibinfo{journal}{\emph{Proceedings of the ACM on Software Engineering}} \bibinfo{volume}{2}, \bibinfo{number}{ISSTA}, \bibinfo{pages}{1054--1076}.
\newblock


\bibitem[Chang and Lin(2016)]%
        {chang2016constraint}
\bibfield{author}{\bibinfo{person}{Cheng-Hung Chang} {and} \bibinfo{person}{Nai-Wei Lin}.} \bibinfo{year}{2016}\natexlab{}.
\newblock \showarticletitle{Constraint-based test case generation for white-box method-level unit testing}. In \bibinfo{booktitle}{\emph{2016 International Computer Symposium (ICS)}}. \bibinfo{pages}{601--604}.
\newblock


\bibitem[Chen et~al\mbox{.}(2021)]%
        {chen2021evaluating}
\bibfield{author}{\bibinfo{person}{Mark Chen}, \bibinfo{person}{Jerry Tworek}, \bibinfo{person}{Heewoo Jun}, \bibinfo{person}{Qiming Yuan}, \bibinfo{person}{Henrique Ponde De~Oliveira Pinto}, \bibinfo{person}{Jared Kaplan}, \bibinfo{person}{Harri Edwards}, \bibinfo{person}{Yuri Burda}, \bibinfo{person}{Nicholas Joseph}, \bibinfo{person}{Greg Brockman}, {et~al\mbox{.}}} \bibinfo{year}{2021}\natexlab{}.
\newblock \showarticletitle{Evaluating large language models trained on code}.
\newblock \bibinfo{journal}{\emph{arXiv preprint arXiv:2107.03374}} (\bibinfo{year}{2021}).
\newblock


\bibitem[Chen et~al\mbox{.}(2024)]%
        {chen2024chatunitest}
\bibfield{author}{\bibinfo{person}{Yinghao Chen}, \bibinfo{person}{Zehao Hu}, \bibinfo{person}{Chen Zhi}, \bibinfo{person}{Junxiao Han}, \bibinfo{person}{Shuiguang Deng}, {and} \bibinfo{person}{Jianwei Yin}.} \bibinfo{year}{2024}\natexlab{}.
\newblock \showarticletitle{Chatunitest: A framework for llm-based test generation}. In \bibinfo{booktitle}{\emph{Companion Proceedings of the 32nd ACM International Conference on the Foundations of Software Engineering}}. \bibinfo{pages}{572--576}.
\newblock


\bibitem[Cheng et~al\mbox{.}(2025)]%
        {Cheng2025}
\bibfield{author}{\bibinfo{person}{Runxiang Cheng}, \bibinfo{person}{Michele Tufano}, \bibinfo{person}{J{\"u}rgen Cito}, \bibinfo{person}{Jos{\'e} Cambronero}, \bibinfo{person}{Pat Rondon}, \bibinfo{person}{Renyao Wei}, \bibinfo{person}{Aaron Sun}, {and} \bibinfo{person}{Satish Chandra}.} \bibinfo{year}{2025}\natexlab{}.
\newblock \showarticletitle{Agentic Bug Reproduction for Effective Automated Program Repair at Google}.
\newblock \bibinfo{journal}{\emph{arXiv preprint arXiv:2502.01821}} (\bibinfo{year}{2025}).
\newblock


\bibitem[Dinella et~al\mbox{.}(2022)]%
        {dinella2022toga}
\bibfield{author}{\bibinfo{person}{Elizabeth Dinella}, \bibinfo{person}{Gabriel Ryan}, \bibinfo{person}{Todd Mytkowicz}, {and} \bibinfo{person}{Shuvendu~K Lahiri}.} \bibinfo{year}{2022}\natexlab{}.
\newblock \showarticletitle{Toga: A neural method for test oracle generation}. In \bibinfo{booktitle}{\emph{Proceedings of the 44th International Conference on Software Engineering}}. \bibinfo{pages}{2130--2141}.
\newblock


\bibitem[Eliseeva et~al\mbox{.}(2025)]%
        {Eliseeva2025}
\bibfield{author}{\bibinfo{person}{Aleksandra Eliseeva}, \bibinfo{person}{Alexander Kovrigin}, \bibinfo{person}{Ilia Kholkin}, \bibinfo{person}{Egor Bogomolov}, {and} \bibinfo{person}{Yaroslav Zharov}.} \bibinfo{year}{2025}\natexlab{}.
\newblock \bibinfo{title}{EnvBench: A Benchmark for Automated Environment Setup}.
\newblock
\showeprint[arxiv]{2503.14443}~[cs.LG]
\urldef\tempurl%
\url{https://arxiv.org/abs/2503.14443}
\showURL{%
\tempurl}


\bibitem[Etemadi et~al\mbox{.}(2024)]%
        {etemadi2024mokav}
\bibfield{author}{\bibinfo{person}{Khashayar Etemadi}, \bibinfo{person}{Bardia Mohammadi}, \bibinfo{person}{Zhendong Su}, {and} \bibinfo{person}{Martin Monperrus}.} \bibinfo{year}{2024}\natexlab{}.
\newblock \showarticletitle{Mokav: Execution-driven differential testing with llms}.
\newblock \bibinfo{journal}{\emph{arXiv preprint arXiv:2406.10375}} (\bibinfo{year}{2024}).
\newblock


\bibitem[Evans and Savoia(2007)]%
        {evans2007differential}
\bibfield{author}{\bibinfo{person}{Robert~B Evans} {and} \bibinfo{person}{Alberto Savoia}.} \bibinfo{year}{2007}\natexlab{}.
\newblock \showarticletitle{Differential testing: a new approach to change detection}. In \bibinfo{booktitle}{\emph{The 6th Joint Meeting on European software engineering conference and the ACM SIGSOFT Symposium on the Foundations of Software Engineering: Companion Papers}}. \bibinfo{pages}{549--552}.
\newblock


\bibitem[Fraser and Arcuri(2011)]%
        {fraser2011evosuite}
\bibfield{author}{\bibinfo{person}{Gordon Fraser} {and} \bibinfo{person}{Andrea Arcuri}.} \bibinfo{year}{2011}\natexlab{}.
\newblock \showarticletitle{Evosuite: automatic test suite generation for object-oriented software}. In \bibinfo{booktitle}{\emph{Proceedings of the 19th ACM SIGSOFT symposium and the 13th European conference on Foundations of software engineering}}. \bibinfo{pages}{416--419}.
\newblock


\bibitem[Hassan et~al\mbox{.}(2018)]%
        {hassan2018studying}
\bibfield{author}{\bibinfo{person}{Safwat Hassan}, \bibinfo{person}{Chakkrit Tantithamthavorn}, \bibinfo{person}{Cor-Paul Bezemer}, {and} \bibinfo{person}{Ahmed~E Hassan}.} \bibinfo{year}{2018}\natexlab{}.
\newblock \showarticletitle{Studying the dialogue between users and developers of free apps in the google play store}.
\newblock \bibinfo{journal}{\emph{Empirical Software Engineering}}  \bibinfo{volume}{23} (\bibinfo{year}{2018}), \bibinfo{pages}{1275--1312}.
\newblock


\bibitem[Jiang et~al\mbox{.}(2025)]%
        {jiang2025cosil}
\bibfield{author}{\bibinfo{person}{Zhonghao Jiang}, \bibinfo{person}{Xiaoxue Ren}, \bibinfo{person}{Meng Yan}, \bibinfo{person}{Wei Jiang}, \bibinfo{person}{Yong Li}, {and} \bibinfo{person}{Zhongxin Liu}.} \bibinfo{year}{2025}\natexlab{}.
\newblock \showarticletitle{CoSIL: Software Issue Localization via LLM-Driven Code Repository Graph Searching}.
\newblock \bibinfo{journal}{\emph{arXiv preprint arXiv:2503.22424}} (\bibinfo{year}{2025}).
\newblock


\bibitem[Jimenez et~al\mbox{.}(2023)]%
        {jimenez2023swe}
\bibfield{author}{\bibinfo{person}{Carlos~E Jimenez}, \bibinfo{person}{John Yang}, \bibinfo{person}{Alexander Wettig}, \bibinfo{person}{Shunyu Yao}, \bibinfo{person}{Kexin Pei}, \bibinfo{person}{Ofir Press}, {and} \bibinfo{person}{Karthik Narasimhan}.} \bibinfo{year}{2023}\natexlab{}.
\newblock \showarticletitle{Swe-bench: Can language models resolve real-world github issues?}
\newblock \bibinfo{journal}{\emph{arXiv preprint arXiv:2310.06770}} (\bibinfo{year}{2023}).
\newblock


\bibitem[Just et~al\mbox{.}(2014)]%
        {just2014defects4j}
\bibfield{author}{\bibinfo{person}{Ren{\'e} Just}, \bibinfo{person}{Darioush Jalali}, {and} \bibinfo{person}{Michael~D Ernst}.} \bibinfo{year}{2014}\natexlab{}.
\newblock \showarticletitle{Defects4J: A database of existing faults to enable controlled testing studies for Java programs}. In \bibinfo{booktitle}{\emph{Proceedings of the 2014 international symposium on software testing and analysis}}. \bibinfo{pages}{437--440}.
\newblock


\bibitem[Kang et~al\mbox{.}(2024)]%
        {kang2024evaluating}
\bibfield{author}{\bibinfo{person}{Sungmin Kang}, \bibinfo{person}{Juyeon Yoon}, \bibinfo{person}{Nargiz Askarbekkyzy}, {and} \bibinfo{person}{Shin Yoo}.} \bibinfo{year}{2024}\natexlab{}.
\newblock \showarticletitle{Evaluating diverse large language models for automatic and general bug reproduction}.
\newblock \bibinfo{journal}{\emph{IEEE Transactions on Software Engineering}} (\bibinfo{year}{2024}).
\newblock


\bibitem[{Le Goues} et~al\mbox{.}(2019)]%
        {cacm2019-program-repair}
\bibfield{author}{\bibinfo{person}{Claire {Le Goues}}, \bibinfo{person}{Michael Pradel}, {and} \bibinfo{person}{Abhik Roychoudhury}.} \bibinfo{year}{2019}\natexlab{}.
\newblock \showarticletitle{Automated program repair}.
\newblock \bibinfo{journal}{\emph{Commun. {ACM}}} \bibinfo{volume}{62}, \bibinfo{number}{12} (\bibinfo{year}{2019}), \bibinfo{pages}{56--65}.
\newblock
\href{https://doi.org/10.1145/3318162}{doi:\nolinkurl{10.1145/3318162}}


\bibitem[Li et~al\mbox{.}(2023)]%
        {li2023nuances}
\bibfield{author}{\bibinfo{person}{Tsz-On Li}, \bibinfo{person}{Wenxi Zong}, \bibinfo{person}{Yibo Wang}, \bibinfo{person}{Haoye Tian}, \bibinfo{person}{Ying Wang}, \bibinfo{person}{Shing-Chi Cheung}, {and} \bibinfo{person}{Jeff Kramer}.} \bibinfo{year}{2023}\natexlab{}.
\newblock \showarticletitle{Nuances are the key: Unlocking chatgpt to find failure-inducing tests with differential prompting}. In \bibinfo{booktitle}{\emph{2023 38th IEEE/ACM International Conference on Automated Software Engineering (ASE)}}. IEEE, \bibinfo{pages}{14--26}.
\newblock


\bibitem[Linares-V{\'a}squez et~al\mbox{.}(2014)]%
        {linares2014api}
\bibfield{author}{\bibinfo{person}{Mario Linares-V{\'a}squez}, \bibinfo{person}{Gabriele Bavota}, \bibinfo{person}{Massimiliano Di~Penta}, \bibinfo{person}{Rocco Oliveto}, {and} \bibinfo{person}{Denys Poshyvanyk}.} \bibinfo{year}{2014}\natexlab{}.
\newblock \showarticletitle{How do api changes trigger stack overflow discussions? a study on the android sdk}. In \bibinfo{booktitle}{\emph{proceedings of the 22nd International Conference on Program Comprehension}}. \bibinfo{pages}{83--94}.
\newblock


\bibitem[Liu et~al\mbox{.}(2024b)]%
        {liu2024your}
\bibfield{author}{\bibinfo{person}{Jiawei Liu}, \bibinfo{person}{Chunqiu~Steven Xia}, \bibinfo{person}{Yuyao Wang}, {and} \bibinfo{person}{Lingming Zhang}.} \bibinfo{year}{2024}\natexlab{b}.
\newblock \showarticletitle{Is your code generated by chatgpt really correct? rigorous evaluation of large language models for code generation}.
\newblock \bibinfo{journal}{\emph{Advances in Neural Information Processing Systems}}  \bibinfo{volume}{36} (\bibinfo{year}{2024}).
\newblock


\bibitem[Liu et~al\mbox{.}(2019)]%
        {liu2019tbar}
\bibfield{author}{\bibinfo{person}{Kui Liu}, \bibinfo{person}{Anil Koyuncu}, \bibinfo{person}{Dongsun Kim}, {and} \bibinfo{person}{Tegawend{\'e}~F Bissyand{\'e}}.} \bibinfo{year}{2019}\natexlab{}.
\newblock \showarticletitle{TBar: Revisiting template-based automated program repair}. In \bibinfo{booktitle}{\emph{Proceedings of the 28th ACM SIGSOFT international symposium on software testing and analysis}}. \bibinfo{pages}{31--42}.
\newblock


\bibitem[Liu et~al\mbox{.}(2021)]%
        {liu2021critical}
\bibfield{author}{\bibinfo{person}{Kui Liu}, \bibinfo{person}{Li Li}, \bibinfo{person}{Anil Koyuncu}, \bibinfo{person}{Dongsun Kim}, \bibinfo{person}{Zhe Liu}, \bibinfo{person}{Jacques Klein}, {and} \bibinfo{person}{Tegawend{\'e}~F Bissyand{\'e}}.} \bibinfo{year}{2021}\natexlab{}.
\newblock \showarticletitle{A critical review on the evaluation of automated program repair systems}.
\newblock \bibinfo{journal}{\emph{Journal of Systems and Software}}  \bibinfo{volume}{171} (\bibinfo{year}{2021}), \bibinfo{pages}{110817}.
\newblock


\bibitem[Liu et~al\mbox{.}(2024a)]%
        {liu2024llm}
\bibfield{author}{\bibinfo{person}{Kaibo Liu}, \bibinfo{person}{Yiyang Liu}, \bibinfo{person}{Zhenpeng Chen}, \bibinfo{person}{Jie~M Zhang}, \bibinfo{person}{Yudong Han}, \bibinfo{person}{Yun Ma}, \bibinfo{person}{Ge Li}, {and} \bibinfo{person}{Gang Huang}.} \bibinfo{year}{2024}\natexlab{a}.
\newblock \showarticletitle{Llm-powered test case generation for detecting tricky bugs}.
\newblock \bibinfo{journal}{\emph{arXiv preprint arXiv:2404.10304}} (\bibinfo{year}{2024}).
\newblock


\bibitem[Lukasczyk and Fraser(2022)]%
        {lukasczyk2022pynguin}
\bibfield{author}{\bibinfo{person}{Stephan Lukasczyk} {and} \bibinfo{person}{Gordon Fraser}.} \bibinfo{year}{2022}\natexlab{}.
\newblock \showarticletitle{Pynguin: Automated unit test generation for python}. In \bibinfo{booktitle}{\emph{Proceedings of the ACM/IEEE 44th International Conference on Software Engineering: Companion Proceedings}}. \bibinfo{pages}{168--172}.
\newblock


\bibitem[Martinez et~al\mbox{.}(2015)]%
        {martinez2015automatic}
\bibfield{author}{\bibinfo{person}{Matias Martinez}, \bibinfo{person}{Thomas Durieux}, \bibinfo{person}{Jifeng Xuan}, \bibinfo{person}{Romain Sommerard}, {and} \bibinfo{person}{Martin Monperrus}.} \bibinfo{year}{2015}\natexlab{}.
\newblock \showarticletitle{Automatic repair of real bugs: An experience report on the defects4j dataset}.
\newblock \bibinfo{journal}{\emph{arXiv preprint arXiv:1505.07002}} (\bibinfo{year}{2015}).
\newblock


\bibitem[Milliken et~al\mbox{.}(2025)]%
        {Milliken2025}
\bibfield{author}{\bibinfo{person}{Louis Milliken}, \bibinfo{person}{Sungmin Kang}, {and} \bibinfo{person}{Shin Yoo}.} \bibinfo{year}{2025}\natexlab{}.
\newblock \showarticletitle{Beyond pip install: Evaluating llm agents for the automated installation of python projects}. In \bibinfo{booktitle}{\emph{2025 IEEE International Conference on Software Analysis, Evolution and Reengineering (SANER)}}. IEEE, \bibinfo{pages}{1--11}.
\newblock


\bibitem[Mündler et~al\mbox{.}(2024)]%
        {Muendler2024}
\bibfield{author}{\bibinfo{person}{Niels Mündler}, \bibinfo{person}{Mark~Niklas Müller}, \bibinfo{person}{Jingxuan He}, {and} \bibinfo{person}{Martin Vechev}.} \bibinfo{year}{2024}\natexlab{}.
\newblock \bibinfo{title}{Code Agents are State of the Art Software Testers}.
\newblock
\showeprint[arxiv]{2406.12952}~[cs.SE]
\urldef\tempurl%
\url{https://arxiv.org/abs/2406.12952}
\showURL{%
\tempurl}


\bibitem[Nashid et~al\mbox{.}(2025)]%
        {Issue2Test_arXiv2025}
\bibfield{author}{\bibinfo{person}{Noor Nashid}, \bibinfo{person}{Islem Bouzenia}, \bibinfo{person}{Michael Pradel}, {and} \bibinfo{person}{Ali Mesbah}.} \bibinfo{year}{2025}\natexlab{}.
\newblock \showarticletitle{Issue2Test: Generating Reproducing Test Cases from Issue Reports}.
\newblock \bibinfo{journal}{\emph{arXiv preprint arXiv:2503.16320}} (\bibinfo{year}{2025}).
\newblock


\bibitem[Pacheco and Ernst(2007)]%
        {pacheco2007randoop}
\bibfield{author}{\bibinfo{person}{Carlos Pacheco} {and} \bibinfo{person}{Michael~D Ernst}.} \bibinfo{year}{2007}\natexlab{}.
\newblock \showarticletitle{Randoop: feedback-directed random testing for Java}. In \bibinfo{booktitle}{\emph{Companion to the 22nd ACM SIGPLAN conference on Object-oriented programming systems and applications companion}}. \bibinfo{pages}{815--816}.
\newblock


\bibitem[Pizzorno and Berger(2024)]%
        {pizzorno2024coverup}
\bibfield{author}{\bibinfo{person}{Juan~Altmayer Pizzorno} {and} \bibinfo{person}{Emery~D Berger}.} \bibinfo{year}{2024}\natexlab{}.
\newblock \showarticletitle{Coverup: Coverage-guided llm-based test generation}.
\newblock \bibinfo{journal}{\emph{arXiv preprint arXiv:2403.16218}} (\bibinfo{year}{2024}).
\newblock


\bibitem[Qi et~al\mbox{.}(2015)]%
        {qi2015analysis}
\bibfield{author}{\bibinfo{person}{Zichao Qi}, \bibinfo{person}{Fan Long}, \bibinfo{person}{Sara Achour}, {and} \bibinfo{person}{Martin Rinard}.} \bibinfo{year}{2015}\natexlab{}.
\newblock \showarticletitle{An analysis of patch plausibility and correctness for generate-and-validate patch generation systems}. In \bibinfo{booktitle}{\emph{Proceedings of the 2015 international symposium on software testing and analysis}}. \bibinfo{pages}{24--36}.
\newblock


\bibitem[Rao et~al\mbox{.}(2023)]%
        {rao2023cat}
\bibfield{author}{\bibinfo{person}{Nikitha Rao}, \bibinfo{person}{Kush Jain}, \bibinfo{person}{Uri Alon}, \bibinfo{person}{Claire Le~Goues}, {and} \bibinfo{person}{Vincent~J Hellendoorn}.} \bibinfo{year}{2023}\natexlab{}.
\newblock \showarticletitle{CAT-LM training language models on aligned code and tests}. In \bibinfo{booktitle}{\emph{2023 38th IEEE/ACM International Conference on Automated Software Engineering (ASE)}}. IEEE, \bibinfo{pages}{409--420}.
\newblock


\bibitem[Roychoudhury et~al\mbox{.}(2025)]%
        {AgenticAISE2025}
\bibfield{author}{\bibinfo{person}{Abhik Roychoudhury}, \bibinfo{person}{Corina Pasareanu}, \bibinfo{person}{Michael Pradel}, {and} \bibinfo{person}{Baishakhi Ray}.} \bibinfo{year}{2025}\natexlab{}.
\newblock \showarticletitle{Agentic AI Software Engineer: Programming with Trust}.
\newblock \bibinfo{journal}{\emph{arXiv preprint arXiv:2502.13767}} (\bibinfo{year}{2025}).
\newblock


\bibitem[Ryan et~al\mbox{.}(2024)]%
        {ryan2024code}
\bibfield{author}{\bibinfo{person}{Gabriel Ryan}, \bibinfo{person}{Siddhartha Jain}, \bibinfo{person}{Mingyue Shang}, \bibinfo{person}{Shiqi Wang}, \bibinfo{person}{Xiaofei Ma}, \bibinfo{person}{Murali~Krishna Ramanathan}, {and} \bibinfo{person}{Baishakhi Ray}.} \bibinfo{year}{2024}\natexlab{}.
\newblock \showarticletitle{Code-aware prompting: A study of coverage-guided test generation in regression setting using llm}.
\newblock \bibinfo{journal}{\emph{Proceedings of the ACM on Software Engineering}} \bibinfo{volume}{1}, \bibinfo{number}{FSE} (\bibinfo{year}{2024}), \bibinfo{pages}{951--971}.
\newblock


\bibitem[Smith et~al\mbox{.}(2015)]%
        {smith2015cure}
\bibfield{author}{\bibinfo{person}{Edward~K Smith}, \bibinfo{person}{Earl~T Barr}, \bibinfo{person}{Claire Le~Goues}, {and} \bibinfo{person}{Yuriy Brun}.} \bibinfo{year}{2015}\natexlab{}.
\newblock \showarticletitle{Is the cure worse than the disease? overfitting in automated program repair}. In \bibinfo{booktitle}{\emph{Proceedings of the 10th Joint Meeting on Foundations of Software Engineering}}. \bibinfo{pages}{532--543}.
\newblock


\bibitem[Su et~al\mbox{.}(2025)]%
        {su2025learn}
\bibfield{author}{\bibinfo{person}{Hongjin Su}, \bibinfo{person}{Ruoxi Sun}, \bibinfo{person}{Jinsung Yoon}, \bibinfo{person}{Pengcheng Yin}, \bibinfo{person}{Tao Yu}, {and} \bibinfo{person}{Sercan~{\"O} Ar{\i}k}.} \bibinfo{year}{2025}\natexlab{}.
\newblock \showarticletitle{Learn-by-interact: A Data-Centric Framework For Self-Adaptive Agents in Realistic Environments}. In \bibinfo{booktitle}{\emph{Proceedings of the Thirteenth International Conference on Learning Representations}}.
\newblock


\bibitem[Tillmann and De~Halleux(2008)]%
        {tillmann2008pex}
\bibfield{author}{\bibinfo{person}{Nikolai Tillmann} {and} \bibinfo{person}{Jonathan De~Halleux}.} \bibinfo{year}{2008}\natexlab{}.
\newblock \showarticletitle{Pex--white box test generation for. net}. In \bibinfo{booktitle}{\emph{International conference on tests and proofs}}. Springer, \bibinfo{pages}{134--153}.
\newblock


\bibitem[Tufano et~al\mbox{.}(2020)]%
        {tufano2020unit}
\bibfield{author}{\bibinfo{person}{Michele Tufano}, \bibinfo{person}{Dawn Drain}, \bibinfo{person}{Alexey Svyatkovskiy}, \bibinfo{person}{Shao~Kun Deng}, {and} \bibinfo{person}{Neel Sundaresan}.} \bibinfo{year}{2020}\natexlab{}.
\newblock \showarticletitle{Unit test case generation with transformers and focal context}.
\newblock \bibinfo{journal}{\emph{arXiv preprint arXiv:2009.05617}} (\bibinfo{year}{2020}).
\newblock


\bibitem[Wang et~al\mbox{.}(2024a)]%
        {wang2024openhands}
\bibfield{author}{\bibinfo{person}{Xingyao Wang}, \bibinfo{person}{Boxuan Li}, \bibinfo{person}{Yufan Song}, \bibinfo{person}{Frank~F Xu}, \bibinfo{person}{Xiangru Tang}, \bibinfo{person}{Mingchen Zhuge}, \bibinfo{person}{Jiayi Pan}, \bibinfo{person}{Yueqi Song}, \bibinfo{person}{Bowen Li}, \bibinfo{person}{Jaskirat Singh}, {et~al\mbox{.}}} \bibinfo{year}{2024}\natexlab{a}.
\newblock \showarticletitle{Openhands: An open platform for ai software developers as generalist agents}.
\newblock \bibinfo{journal}{\emph{arXiv preprint arXiv:2407.16741}} (\bibinfo{year}{2024}).
\newblock


\bibitem[Wang et~al\mbox{.}(2024b)]%
        {wang2024hits}
\bibfield{author}{\bibinfo{person}{Zejun Wang}, \bibinfo{person}{Kaibo Liu}, \bibinfo{person}{Ge Li}, {and} \bibinfo{person}{Zhi Jin}.} \bibinfo{year}{2024}\natexlab{b}.
\newblock \showarticletitle{HITS: High-coverage LLM-based Unit Test Generation via Method Slicing}. In \bibinfo{booktitle}{\emph{Proceedings of the 39th IEEE/ACM International Conference on Automated Software Engineering}}. \bibinfo{pages}{1258--1268}.
\newblock


\bibitem[Xia et~al\mbox{.}(2025)]%
        {xia2024agentless}
\bibfield{author}{\bibinfo{person}{Chunqiu~Steven Xia}, \bibinfo{person}{Yinlin Deng}, \bibinfo{person}{Soren Dunn}, {and} \bibinfo{person}{Lingming Zhang}.} \bibinfo{year}{2025}\natexlab{}.
\newblock \showarticletitle{Agentless: Demystifying llm-based software engineering agents}. In \bibinfo{booktitle}{\emph{Proceedings of 33rd ACM SIGSOFT International Symposium on the Foundations of Software Engineering}}.
\newblock


\bibitem[Xin and Reiss(2017)]%
        {xin2017identifying}
\bibfield{author}{\bibinfo{person}{Qi Xin} {and} \bibinfo{person}{Steven~P Reiss}.} \bibinfo{year}{2017}\natexlab{}.
\newblock \showarticletitle{Identifying test-suite-overfitted patches through test case generation}. In \bibinfo{booktitle}{\emph{Proceedings of the 26th ACM SIGSOFT international symposium on software testing and analysis}}. \bibinfo{pages}{226--236}.
\newblock


\bibitem[Xiong et~al\mbox{.}(2018)]%
        {xiong2018identifying}
\bibfield{author}{\bibinfo{person}{Yingfei Xiong}, \bibinfo{person}{Xinyuan Liu}, \bibinfo{person}{Muhan Zeng}, \bibinfo{person}{Lu Zhang}, {and} \bibinfo{person}{Gang Huang}.} \bibinfo{year}{2018}\natexlab{}.
\newblock \showarticletitle{Identifying patch correctness in test-based program repair}. In \bibinfo{booktitle}{\emph{Proceedings of the 40th international conference on software engineering}}. \bibinfo{pages}{789--799}.
\newblock


\bibitem[Xu et~al\mbox{.}(2025)]%
        {xu2025flexfl}
\bibfield{author}{\bibinfo{person}{Chuyang Xu}, \bibinfo{person}{Zhongxin Liu}, \bibinfo{person}{Xiaoxue Ren}, \bibinfo{person}{Gehao Zhang}, \bibinfo{person}{Ming Liang}, {and} \bibinfo{person}{David Lo}.} \bibinfo{year}{2025}\natexlab{}.
\newblock \showarticletitle{Flexfl: Flexible and effective fault localization with open-source large language models}.
\newblock \bibinfo{journal}{\emph{IEEE Transactions on Software Engineering}} (\bibinfo{year}{2025}).
\newblock


\bibitem[Yang et~al\mbox{.}(2024)]%
        {Yang2024a}
\bibfield{author}{\bibinfo{person}{John Yang}, \bibinfo{person}{Carlos~E. Jimenez}, \bibinfo{person}{Alexander Wettig}, \bibinfo{person}{Kilian Lieret}, \bibinfo{person}{Shunyu Yao}, \bibinfo{person}{Karthik Narasimhan}, {and} \bibinfo{person}{Ofir Press}.} \bibinfo{year}{2024}\natexlab{}.
\newblock \showarticletitle{SWE-agent: Agent-Computer Interfaces Enable Automated Software Engineering}. In \bibinfo{booktitle}{\emph{Advances in Neural Information Processing Systems 38: Annual Conference on Neural Information Processing Systems 2024, NeurIPS 2024, Vancouver, BC, Canada, December 10 - 15, 2024}}, \bibfield{editor}{\bibinfo{person}{Amir Globersons}, \bibinfo{person}{Lester Mackey}, \bibinfo{person}{Danielle Belgrave}, \bibinfo{person}{Angela Fan}, \bibinfo{person}{Ulrich Paquet}, \bibinfo{person}{Jakub~M. Tomczak}, {and} \bibinfo{person}{Cheng Zhang}} (Eds.).
\newblock
\urldef\tempurl%
\url{http://papers.nips.cc/paper\_files/paper/2024/hash/5a7c947568c1b1328ccc5230172e1e7c-Abstract-Conference.html}
\showURL{%
\tempurl}


\bibitem[Yang et~al\mbox{.}(2023)]%
        {yang2023large}
\bibfield{author}{\bibinfo{person}{Jun Yang}, \bibinfo{person}{Yuehan Wang}, \bibinfo{person}{Yiling Lou}, \bibinfo{person}{Ming Wen}, {and} \bibinfo{person}{Lingming Zhang}.} \bibinfo{year}{2023}\natexlab{}.
\newblock \showarticletitle{A large-scale empirical review of patch correctness checking approaches}. In \bibinfo{booktitle}{\emph{Proceedings of the 31st ACM Joint European Software Engineering Conference and Symposium on the Foundations of Software Engineering}}. \bibinfo{pages}{1203--1215}.
\newblock


\bibitem[Yu et~al\mbox{.}(2019)]%
        {yu2019alleviating}
\bibfield{author}{\bibinfo{person}{Zhongxing Yu}, \bibinfo{person}{Matias Martinez}, \bibinfo{person}{Benjamin Danglot}, \bibinfo{person}{Thomas Durieux}, {and} \bibinfo{person}{Martin Monperrus}.} \bibinfo{year}{2019}\natexlab{}.
\newblock \showarticletitle{Alleviating patch overfitting with automatic test generation: a study of feasibility and effectiveness for the nopol repair system}.
\newblock \bibinfo{journal}{\emph{Empirical Software Engineering}}  \bibinfo{volume}{24} (\bibinfo{year}{2019}), \bibinfo{pages}{33--67}.
\newblock


\bibitem[Zhang et~al\mbox{.}(2024)]%
        {zhang2024autocoderover}
\bibfield{author}{\bibinfo{person}{Yuntong Zhang}, \bibinfo{person}{Haifeng Ruan}, \bibinfo{person}{Zhiyu Fan}, {and} \bibinfo{person}{Abhik Roychoudhury}.} \bibinfo{year}{2024}\natexlab{}.
\newblock \showarticletitle{Autocoderover: Autonomous program improvement}. In \bibinfo{booktitle}{\emph{Proceedings of the 33rd ACM SIGSOFT International Symposium on Software Testing and Analysis}}. \bibinfo{pages}{1592--1604}.
\newblock


\end{thebibliography}
